\documentstyle[12pt,epsf]{article}

\font\mybb=msbm10 at 12pt
\def\bb#1{\hbox{\mybb#1}}

\begin{document}
\begin{titlepage}
\begin{flushright}
KUNS-1444 \\
HE(TH)97/07 \\
hep-ph/9705449
\end{flushright}

\begin{center}
  \vspace*{1.2cm}

  {\LARGE\bf Finite SUSY GUT Revisited }
  \vspace{1.5cm}

  {\large K{\sc oichi} Y{\sc oshioka} \footnote{e-mail address: {\tt
        yoshioka@gauge.scphys.kyoto-u.ac.jp}}}
  \vspace{0.2cm}

  {\it Department of Physics, Kyoto University \\ Kyoto 606-01, Japan }
  \vspace{1.5cm}

  \begin{abstract}
    We analyze finite supersymmetric SU(5) GUT taking into account the
    problem in the Higgs potential and threshold corrections to gauge
    and Yukawa couplings at GUT and SUSY scales which is important in
    case of large $\tan \beta$ model. We find that even with these
    finiteness conditions, which are very restrictive, there are
    parameter regions where low-energy experimental values are
    consistently reproduced and the Higgs potential actually realizes
    both constraints for large $\tan \beta$ and radiative electroweak
    symmetry breaking, provided that a free parameter is introduced in
    the boundary condition of the Higgs mixing mass parameter.
  \end{abstract}
\end{center}
\end{titlepage}

\newpage

\section{Introduction}
\setcounter{equation}{0}\setcounter{footnote}{0}

The appearance of infinity in quantum field theory has long been one
of the annoying problems. Most of particle physicists would believe
that the ultimate theory, if it exits, should not contain any infinity
and needs no renormalization procedure. In the '80s, it was pointed
out that the requirement of no quadratic divergence leads a kind of
symmetry, supersymmetry \cite{2div}. Then it is interesting to
consider what symmetry appear due to the requirement of vanishing even
logarithmic divergence (vanishing $\beta$-functions) in supersymmetric
theories. Among these theories, $N=4$ and some $N=2$ theories have
zero $\beta$-functions, what is called finiteness, in all-order of
perturbation theory \cite{finitesusy} and they are believed that there
are duality symmetries in those theories \cite{duality}.  In this way,
imposing that there are no infinities in theories corresponds to very
important symmetries until now.  If that is the case, what happens in
$N=1$ theory from the requirement of finiteness?  In perturbative
region, there is the classification table of models in which gauge and
Yukawa couplings satisfy the conditions of finiteness in 1-loop order
\cite{table}. Moreover the all-order finiteness condition for these
couplings is also found \cite{all}. $\,\beta = 0\,$ is also strongly
related to the non-perturbative dynamics such as the electro-magnetic
duality transformation proposed by Seiberg \cite{seiberg}. However,
what (symmetry) corresponds to these finiteness conditions has not
been answered yet.

On the other hand from a viewpoint of low-energy phenomenology, $N=1$
supersymmetry which would come from the requirement of vanishing
quadratic divergences has provided us with many interesting
consequences. Therefore it will be surely important and become a first
step toward understanding of the meanings of finiteness to analyze
phenomenological results as a consequence of vanishing logarithmic
divergence.  The finiteness conditions stated in section 2 prohibit us
from applying them to U(1) gauge theory, and to MSSM.  Then we
apply these conditions to grand unified theory (GUT) and derive the
boundary conditions at GUT scale for couplings of low-energy theory
which we suppose to be MSSM. In particular in case of SU(5)
GUT models, many articles have obtained interesting results of the
fermion masses, superparticle masses, etc.\ \cite{fut,fut2,afut}.

In this paper we analyze this SU(5) model taking care of the following
two points which have not been included explicitly so far.
(1) Since we use the 2-loop order $\beta$-functions for
MSSM couplings we need to include the 1-loop order threshold
corrections \cite{weinberg}.  Especially, it is important in large
$\tan \beta$ model to include a SUSY threshold correction to bottom
quark mass $m_b$ which can be $20 \!\sim\! 40\,$\% of the uncorrected
value \cite{bot}.  (2) Furthermore in large $\tan \beta$ model, we
must check whether the Higgs potential can really generate the
radiative electroweak symmetry breaking \cite{sb} with the large value
of $\tan \beta$. To carry out this, it is necessary to take account of
a new parameter to the boundary condition for the Higgs mixing mass
parameter.

We briefly review the finiteness conditions in $N=1$ supersymmetric
gauge theory and its application the finite SU(5) model in section
2. In section 3 we consider the matching conditions of this finite
SU(5) model to MSSM and calculate low-energy predictions. The GUT and
SUSY threshold corrections which are characteristic to this model are
discussed in section 4. Section 5 is devoted to summary and
comments. The appendices contain the explicit tree-level form of
mass formulae and the SUSY threshold corrections to gauge couplings in
MSSM.

\vspace*{3mm}

\section{Finite SU(5) model}
\setcounter{equation}{0}\setcounter{footnote}{0}

First we describe the all-order finiteness conditions for gauge
couplings and the couplings in superpotential sector.
We consider an anomaly free $N=1$ supersymmetric gauge theory based on
a simple gauge group $G$ with a gauge coupling $g$, and with the
superpotential

\begin{equation}
  W = \frac{1}{2} m_{ij}\Phi^i\Phi^j +
  \frac{1}{6}Y_{ijk}\Phi^i\Phi^j\Phi^k\,.
  \label{W}
\end{equation}
The 1-loop $\beta$-functions are given by
\begin{eqnarray}
  \beta_g^{\,(1)} &\!\! = \!\!& \frac{-g^3}{16\pi^2}\left( 3C_2(G) -
    \sum T(R_i) \right)\,, \\[2mm]
  \beta_{ij}^{\,(1)} &\!\! = \!\!& m_{ik}\gamma^{k\,(1)}_j +
  m_{jk}\gamma^{k\,(1)}_i\,, \\[2mm]
  \beta_{ijk}^{\,(1)} &\!\! = \!\!& Y_{ijl}\gamma^{l\,(1)}_k +
  Y_{jkl}\gamma^{l\,(1)}_i + Y_{kil}\gamma^{l\,(1)}_j\,, \\[3mm]
  && \hspace*{-5mm} \gamma^{i\,(1)}_j = \frac{1}{32\pi^2}\left (
    Y^{ikl} Y_{jkl}-4g^2C_2(R_i) \,\delta^i_j \right) \qquad
  ( Y^{ijk}\equiv Y_{ijk}^*)\,,
\end{eqnarray}
where $\gamma^{i\,(1)}_j$ are the 1-loop anomalous dimensions of the
field $\Phi^i$ and
\begin{eqnarray}
  \mbox{tr} (\,T^a T^b) = T(R)\delta^{ab}\,,\quad \sum_a \,T^aT^a =
  C_2(R)\underline{1} \;,\quad \sum_{c\,,\,d} f_{acd}f_{bcd} =
  C_2(G)\delta_{ab}\,,
\end{eqnarray}
$f_{abc}$ are the structure constants of the gauge group $G$.
Then the necessary and sufficient conditions for 1-loop
order finiteness\footnote{Interestingly enough, these conditions are
  necessary and sufficient
  for 2-loop order finiteness \cite{1to2}. }
(to contain no divergence) are $\beta_g^{\,(1)} = \gamma^{i\,(1)}_j =
0\,$;
\begin{eqnarray}
  3C_2(G) - \sum T(R_i) &\!\! = \!\!& 0 \,, \label{fc1-1} \\[2mm]
  Y^{ikl} Y_{jkl}-4g^2C_2(R_i) \,\delta^i_j  &\!\! = \!\!& 0\,.
  \label{fc1-2}
\end{eqnarray}
That is, the field contents satisfy the condition (\ref{fc1-1}) and
the conditions (\ref{fc1-2}) possess ($g$-expansion) solutions of the
form
\begin{eqnarray}
  \hspace*{1cm}Y_{ijk} = \rho_{ijk}\, g \,,\qquad\;\; \rho_{ijk} \in
  \bb{C}\;(\mbox{const.})\,.
\end{eqnarray}
The field contents which satisfy these conditions are listed in
Ref.$\,$\cite{table}.
In addition to the above conditions it is necessary to impose one more
condition so that this theory may have no divergence in all-order of
perturbation theory. This condition says that the solutions of
vanishing one-loop anomalous dimensions, $\gamma^{i\,(1)}_j = 0$, are
isolated and non-degenerate\footnote{By ``isolated'' and
  ``non-degenerate'', we mean that the solutions  cannot be multiple
  zeroes and are not parameterized either. }
when considered as solutions of vanishing one-loop Yukawa
$\beta$-functions, $\beta_{ijk}^{\,(1)} = 0$ \cite{all}.
Surprisingly enough, the conditions for all-order finiteness can be
expressed in term of the 1-loop order quantities ($\beta$-functions).
Therefore one may easily apply these conditions to definite models.

Next we discuss soft SUSY breaking sector. In general, this part of
potential takes the form (for simple gauge group $G$),
\begin{eqnarray}
  V_{soft} = {1\over 2} (\mu^2)_j^i \phi^*_i \phi^j + {1\over 2}
  B_{ij} \phi^i \phi^j + {1\over 6} h_{ijk} \phi^i \phi^j \phi^k +
  {1\over 2} M \lambda \lambda + {\rm h.c.}
\end{eqnarray}
where here and hereafter the fields $\phi^i$ in $V_{soft}$ denote the
scalar component of those superfields and $\lambda$ means gaugino.
For these parameters in $V_{soft}$,\ the (renormalization-group
invariant) finiteness conditions are found within only 2-loop order
\cite{soft} unlike for the dimensionless couplings.  These conditions
are given by
\begin{eqnarray}
  h_{ijk} &\!\! = \!\!& - M \,Y_{ijk}\,, \label{fc2-1} \\[2mm]
  (\mu^2)^i_j &\!\! = \!\!& \displaystyle\frac{1}{3} \,M
  M^*\delta^i_j\,,
  \label{fc2-2}
\end{eqnarray}
together with (\ref{fc1-1}), (\ref{fc1-2}).\footnote{The
  renormalization-group invariant relation for $B^{ij}$ \cite{soft} is
  not required by finiteness (within 2-loop level). } \
With these conditions, the theory has $\beta$-functions that all
vanish at least to 2-loop order. It is interesting to note that these
universal forms of finiteness condition are the same as those derived
from superstring or $N=1$ supergravity models \cite{soft}.

As a concrete example of phenomenological applications of the
finiteness conditions, we consider supersymmetric SU(5) GUT
models.  According to the classification tables in
Ref.$\,$\cite{table}, \ there exit only one field content which
fulfills the following requirements;
\vspace*{-1mm}
\begin{list}{$\cdot$}{}
\item It contains chiral three families (three $( {\bf {\bar
      5}}\,,\,{\bf 10} )$ sets). \vspace*{-3mm}
\item The remains of the contents are vector-like ones.
  \vspace*{-3mm}
\item It contains fields in an adjoint representation to break
  the GUT gauge group.
\end{list}
\vspace*{-1mm}
This model contains $( \bf
5\,,\,\bar{5}\,,\,10\,,\,\overline{10}\,,\,24  )$
with the multiplicities $( 4\,,\,7\,,\,3\,,\,0\,,\,1 )$. Then the
general superpotential for this contents is
\begin{eqnarray}
  W &\!\! = \!\!& {1\over 2} f_{ija} 10_i 10 _j H_a + {\bar f}_{ija}
  10_i {\bar 5}_j {\bar H}_a +{1\over 2} q_{ijk} 10_i
  {\bar 5}_j {\bar 5}_k \\[2mm]
  &&\hspace*{0.1cm} + {1\over 2} q^\prime_{ia b}
  10_i {\bar H}_a {\bar H}_b +f_{a b} {\bar H}_a \Sigma H_b + p
  \,\Sigma^3 + f_{i a} {\bar 5}_i \Sigma H_a + W_m  \nonumber
\end{eqnarray}
\begin{center}
  $(i,j = 1, 2, 3 \,,\quad a,b = 1, \cdots ,4)$
\end{center}
where $W_m$ contains the mass terms of $H_a\,,\,{\bar
  H}_a\,,\,\Sigma$. \
It is sufficient to impose following discrete symmetries in order to
get isolate and non-degenerate solutions and suppress the rapid
nucleon decay \cite{afut}.
\begin{center}
\begin{tabular}{|c|c|c|c|c|c|c|c|c|c|c|c|}
  \hline & $10_1$ & $10_2$ & $10_3$ & $\bar 5_1$ & $\bar 5_2$ & $\bar
  5_3$& $H_1$ & $H_2$ & $H_3$ & $H_4$ & $\Sigma$  \\ \hline
  $Z_7$ & 1 & 2 & 4 & 4 & 1 & 2 & 5 & 3 & 6 & 0 & 0 \\ \hline
  $Z_3$ & 1 & 2 & 0 & 0 & 0 & 0 & 1 & 2 & 0 & 0 & 0 \\ \hline
  $Z_2$ & 1 & 1 & 1 & 1 & 1 & 1 & 0 & 0 & 0 & 0 & 0 \\ \hline
\end{tabular}\\
\vspace*{0.5cm}
  Table 1 : The charges of the $Z_7\times Z_3\times Z_2$ (matter
  parity) \vspace*{0.1cm}

($\overline{H}_a$ has an opposite charge to that of $H_a$.)
\end{center}
With these symmetries, the superpotential is restricted to the form,
\begin{eqnarray}
  W = {1\over 2} f_{iii} 10_i 10 _i H_i + {\bar
    f}_{iii} 10_i {\bar 5}_i {\bar H}_i +f_{aa} {\bar H}_a
  \Sigma H_a + p \,\Sigma^3 + W_m
\label{sp}
\end{eqnarray}
and the unique solution which guarantees the all-order finiteness are
given by \cite{fut,fut2,afut}
\begin{eqnarray}
\begin{array}{c}
  f_{111} = f_{222} = f_{333} = \displaystyle{\sqrt{8\over 5}}\, g
  \,,~~~~~ {\bar f}_{111} = {\bar f}_{222} = {\bar
    f}_{333} = \displaystyle{\sqrt{6\over 5}}\, g \,,\\[4mm]
  \!\!\!\!f_{11} = f_{22} = f_{33} = 0 \,,~~~~~f_{44} = g\,,~~~~~p =
  \displaystyle{\sqrt{15\over 7}}\, g\,.
\end{array}
\label{sol1}
\end{eqnarray}

Then we apply the finiteness conditions to $V_{soft}$ to get final
expression of this SU(5) model. The general potential takes the form
\begin{eqnarray}
  V_{soft} &\!\! = \!\!& (m^2_5)_{ij} \bar{5}^\dagger_i \bar{5}_j +
  (m^2_{10})_{ij} 10^\dagger_i 10_j + (m^2_H)_{ab} H_a^\dagger H_b +
  (m^2_{\bar H})_{ab} \bar{H}_a^\dagger \bar{H}_b \nonumber \\[3mm]
  && \quad +\, m^2_{\Sigma} \Sigma^\dagger \Sigma + \biggl
  ( \;\frac{1}{2} M \lambda \lambda + B_{ab} \bar{H}_a H_b +
  B_{\Sigma} \Sigma \Sigma \label{pot} \\[2mm] &&
  \qquad  + \frac{h_{ijk}}{2} 10_i 10_j H_k + {\bar h}_{ijk} 10_i
    {\bar 5}_j {\bar H}_k + h_{ab} {\bar H}_a \Sigma H_b + h_p
    \,\Sigma^3 + \mbox{h.c.} \biggr)\,. \nonumber
\end{eqnarray}
Taking into account the form of the superpotential (\ref{sp}), we can
get the relations among the soft parameters for 2-loop order
finiteness from (\ref{fc2-1}) and (\ref{fc2-2})$\,$;
\begin{eqnarray}
  \begin{array}{c}
    h_{iii} = - M f_{iii} \,,\quad {\bar h}_{iii} = - M
    \bar{f}_{iii}\,,\quad h_{44} = - M f_{44} \,,\quad h_p = - M p\,,
    \\[5mm]
    (m^2_H)_{ab} = (m^2_{\bar H})_{ab} = \displaystyle{\frac{1}{3}}
    M^2 \delta_{ab}\,,\quad m^2_{\Sigma} = \displaystyle{\frac{1}{3}}
    M^2\,,\quad (m^2_{10})_{ij} = (m^2_5)_{ij} =
    \displaystyle{\frac{1}{3}} M^2 \delta_{ij}\,,
  \end{array}
  \label{sol2}
\end{eqnarray}
and all other elements are zero. This, together with the relations
(\ref{sol1}), provides us with the finite\footnote{All-order finite
  for gauge and Yukawa couplings, whereas (at least) 2-loop order
  finite for soft SUSY breaking parameters. } SU(5) model above GUT
scale.
Now, note that from the conditions of finiteness there are no
constraints for the B-parameters in the potential (\ref{pot}) as well
as the supersymmetric Higgs mass parameters.  These are to be
determined by the requirements of low-energy assumptions, which is the
task of the next section.

\section{Matching to MSSM and low-energy predictions}
\setcounter{equation}{0}\setcounter{footnote}{0}

In this section, we analyze low-energy predictions of the finite
SU(5) model. This model which is supposed to break spontaneously to
MSSM at GUT scale $M_G$ casts the boundary conditions for the
couplings of MSSM from the finiteness conditions. Leaving GUT
threshold corrections to gauge and Yukawa couplings in the next
section, we first consider the matching of the parameters between MSSM
and the finite SU(5) at tree level.

The matching conditions for gauge couplings are trivial,
\begin{eqnarray}
  g_1(M_G) = g_2(M_G) = g_3(M_G) = g \,.
\label{gau}
\end{eqnarray}
where $g_1\,,\,g_2\,,\,g_3$ are gauge couplings of
$U(1)_Y\,,\,SU(2)_W\,,\,SU(3)_C$.

As for Yukawa couplings, one may think that from the solutions
(\ref{sol1}) a pair of light Higgs doublet which result from the
doublet-triplet splitting mechanism don't couple to any of the matter
fields $({\bf {\bar 5}\,,\,10})$.
However as mentioned in the end of section 2, the Higgs mass
parameters in $W_m$ are not constrained from finiteness and have a
freedom for tuning of mass parameters of $H \bar H$ to cause a Higgs
mixing at $M_G$. In this paper we take the nonzero couplings
of the Higgs doublets to only the third generation matter fields but
the Yukawa couplings for the first and second
generations are obtained in the same way \cite{softfut}.

After SU(5) symmetry breaking, we suppose the supersymmetric $H \bar
H$ mass terms take the form,
\begin{eqnarray}
  W'_m  &\!\! = \!\!& f_{44}\bar{H}_4 \left\langle\,
    \Sigma\,\right\rangle H_4 + M_1
  \bar{H}_1 H_1 + M_2 \bar{H}_2 H_2 + M_{ab} \bar{H}_a H_b
  \,,\quad (a,b = 3,4) \label{m12} \\[2mm]
  && \qquad\qquad\qquad M_1\,,\;M_2 \sim M_G \,. \nonumber
\end{eqnarray}
Substituting $\left\langle\, \Sigma\,\right\rangle = \omega\cdot
\mbox{diag}(2,2,2,-3,-3)$ into $W_m$,
the mass terms of $H_3$ and $H_4$ become as follows;
\begin{eqnarray}
  && \bar{H}_a^{(3)}M_{ab}^{(3)}H_b^{(3)} \equiv \bar{H}_a^{(3)}\left
    ( M_{ab} + \pmatrix{0& \cr &2\cr}\omega f_{44}
  \right)H_b^{(3)}\,,\\[2mm]
  &&\bar{H}_a^{(2)}M_{ab}^{(2)}H_b^{(2)} \equiv \bar{H}_a^{(2)}\left
    ( M_{ab} + \pmatrix{0& \cr &-3\cr}\omega f_{44}
  \right)H_b^{(2)}\,, \label{h2m} \\[2mm]
  && \quad\qquad H = \pmatrix{H^{(3)} \cr H^{(2)} \cr}\,, \qquad
  \bar{H} = \pmatrix{\bar{H}^{(3)} \cr \bar{H}^{(2)} \cr}\,.
\end{eqnarray}
We diagonalize $M^{(2)}$ of the pair of light Higgs doublets
\begin{eqnarray}
  &&\!\!H'_a = \pmatrix{\cos \theta & \sin \theta \cr -\sin \theta &
    \cos \theta \cr}H_a \,,\quad \bar{H}'_a = \pmatrix{\cos
    \bar{\theta} & \sin \bar{\theta} \cr -\sin \bar{\theta} & \cos
    \bar{\theta} \cr}\bar{H}_a \,,
\end{eqnarray}
\begin{eqnarray}
  &&\hspace*{-5mm}M^{(2)'} = \pmatrix{\cos \bar{\theta} & \sin
    \bar{\theta} \cr -\sin \bar{\theta} & \cos \bar{\theta}
    \cr}M^{(2)}\pmatrix{\cos \theta & -\sin \theta \cr \sin \theta &
    \cos \theta \cr} \equiv \pmatrix{\mu&0 \cr 0&\mu' \cr} \,,
  \label{h2m'} \\[3mm]
  &&\quad \qquad\qquad \mu \ll M_G  \; ( \sim
  M_W )\,,\quad  \mu' \sim  M_G
\end{eqnarray}
which leads to the triplet Higgs mass terms$\,$;
\begin{eqnarray}
  &&\hspace*{-1cm}M^{(3)'} = \pmatrix{\mu&0 \cr 0&\mu' \cr} + 5 \omega
  f_{44} \pmatrix{\sin \theta \sin \bar{\theta} &\cos \theta \sin
    \bar{\theta} \cr \sin \theta \cos \bar{\theta} & \cos \theta \cos
    \bar{\theta} \cr} \,.
\label{h3m'}
\end{eqnarray}
After this rotation, the Yukawa couplings of the third generations to
the light Higgs become
\begin{eqnarray}
  f_{333}' = f_{333}\cos \theta \,,\qquad \bar{f}_{333}' =
  \bar{f}_{333} \cos \bar{\theta} \,.
\label{rot}
\end{eqnarray}
{}From this, we can consider the following two separate cases:
\begin{list}{$\cdot$}{}
\item case (1) : \ $\cos \theta = \cos \bar{\theta} \,\sim\, 1$ \ \ \
  ( large $\tan\beta$ ). \vspace*{-2mm}
\item case (2) : \ $\cos \theta \sim 1\,,\, \cos \bar{\theta} \sim 0$
  \ ( small $\tan\beta$ ). \vspace*{-1mm}
\end{list}
In this paper, we use the next values for each case; (1) $\cos \theta
= \cos \bar{\theta} = 0.9856 \;( \sin \theta = \sin \bar{\theta} =
0.169 )$ and (2) $\cos \theta = 0.954\,,\,\cos \bar{\theta} = 0.03 \;
( \sin \theta = 0.3\,,\,\sin \bar{\theta} = 0.9995 )$.\footnote{For
  $\mu \sim 100$ GeV, $\mu' \sim 10^{16}$ GeV, we can take the mass
  parameters: \ (1) $M_{33} \sim 0.6\,,\,M_{34} = M_{43} \sim
  3.4\,,\,M_{44} \sim 20\,$ and (2) $M_{33} \sim 3.0\,,\,M_{34} \sim
  9.5\,,\,M_{43} \sim 0.1\,,\,M_{44} \sim 0.3\;\;(\times 10^{16}\,
  \mbox{GeV}\,)$. }
\ With the above assumptions the superpotential in MSSM take the form,
\begin{equation}
  W = y_t H Q_3\overline{t} + y_b \bar{H} Q_3\overline{b} +
  y_{\tau} \bar{H} L_3\overline{\tau} + \rho \bar{H} H \,.
\end{equation}
In the end, for these couplings the matching conditions at $M_G$ turn
out,
\begin{eqnarray}
  y_t (M_G) = f'_{333} = f_{333} \cos \theta \,,&\!\!\!\!\!& y_b(M_G)
  = y_\tau(M_G) = \bar{f}'_{333} = \bar{f}_{333} \cos \bar{\theta}\,,
  \label{yukawa} \\[3mm]
  &&\hspace*{-6mm}\rho \, (M_G) = \mu\,. \label{mu}
\end{eqnarray}

Finally, we consider the matching conditions in the soft SUSY breaking
sector.  In MSSM, a general form for this sector is
\begin{eqnarray}
  V_{soft} &\!\! = \!\!& m_1^2 \bar{H}^{\dagger}\bar{H} + m_2^2
  H^{\dagger}H + (m_3^2 \bar{H} H + \mbox{h.c.} ) \nonumber \\[3mm]
  && + \sum_{i=1,2,3} \left( m_{\tilde Q_i}^2 |Q_i|^2 + m_{\tilde
      L_i}^2 |L_i|^2 + m^2_{\tilde{u}_i} |\bar{u}_i |^2
    +m^2_{\tilde{d}_i} |\bar{d}_i |^2 + m^2_{\tilde{e}_i} |\bar{e}_i
    |^2\right)\nonumber \\[1.5mm]
  && \qquad + \left( h_t H Q_3\bar{t} + h_b \bar{H} Q_3\bar{b} +
    h_{\tau} \bar{H} L_3\bar{\tau} + \mbox{h.c.} \right) \\[2mm]
  && \qquad + \frac{1}{2} \Bigl( M_{\lambda_1} \lambda_1\lambda_1 +
  M_{\lambda_2} \lambda_2\lambda_2 + M_{\lambda_3} \lambda_3\lambda_3
  + \mbox{h.c.} \Bigr)\,. \nonumber
\end{eqnarray}
Taking into account the above Higgs rotation, the matching conditions
for these parameters (except for $m_3^2\,$) become as follows;
\begin{eqnarray}
  &&\hspace*{-8mm} h_t(M_G) = h_{333}\cos \theta \,,\quad h_b(M_G) =
  h_\tau(M_G) = \bar{h}_{333} \cos \bar{\theta}\,, \label{h}\\[2mm]
  &&\hspace*{0.2cm} m_{\tilde Q_i}^2(M_G) = m^2_{\tilde{u}_i}(M_G) =
  m^2_{\tilde{e}_i}(M_G)  = m^2_{10\,i} \,, \\[2mm]
  && \hspace*{1.2cm} m_{\tilde L_i}^2(M_G) = m^2_{\tilde{d}_i}(M_G) =
  m^2_{5\,i} \\[2mm]
  &&\hspace*{-0.1cm} m_1^2(M_G) = (m^2_{\bar{H}'})_{33}\,,\qquad
  m_2^2(M_G) = (m^2_{H'})_{33}\,, \\[2mm]
  &&\hspace*{0.3cm}M_{\lambda_1}(M_G) = M_{\lambda_2}(M_G) =
  M_{\lambda_3}(M_G) = M \,. \label{m}
\end{eqnarray}
As a consequence of finiteness, from (\ref{sol1}), (\ref{sol2}),
(\ref{gau}), (\ref{yukawa}), (\ref{mu}) and (\ref{h}) $\sim$ (\ref{m})
we obtain the boundary conditions for MSSM couplings at $M_G$,
\begin{eqnarray}
  &&\hspace*{2.5cm}g_1 = g_2 = g_3 = g\,, \label{bcg}\\[1mm]
  &&\hspace*{0.3cm} y_t = \sqrt{\frac{8}{5}}\, g \cos \theta \,,\qquad
  y_b = y_\tau = \sqrt{\frac{6}{5}}\, g \cos \bar{\theta}\,,
  \label{bcy}\\[1mm]
  &&\hspace*{-0.1cm}h_t = - \sqrt{\frac{8}{5}} M g \cos
  \theta\,,\qquad h_b = h_\tau = - \sqrt{\frac{6}{5}} M g \cos
  \bar{\theta}\,, \\[1mm]
  &&\hspace*{-1.5cm}m_{\tilde Q_i}^2 = m_{\tilde L_i}^2 =
  m^2_{\tilde{u}_i} = m^2_{\tilde{d}_i} = m^2_{\tilde{e}_i} = m_1^2 =
  m_2^2 = \frac{1}{3} M^2\,,\qquad (i = 1,2,3) \\[3mm]
  &&\hspace*{1.8cm}M_{\lambda_1} = M_{\lambda_2} = M_{\lambda_3} =
  M\,, \\[1mm]
  &&\hspace*{3.4cm}\rho = \mu \label{bcm}\,.
\end{eqnarray}

Furthermore we can determine a boundary condition for $m_3^2$ when we
consider to cause the doublet-triplet splitting in the soft SUSY
breaking sector as well as in $W'_m$. After SU(5) breaking, the soft
mass terms of $H^{(2)}_a$ become from (\ref{sol2}),
\begin{eqnarray}
  V_{soft}^{H^{(2)}} \,= \,\bar{H}_a^{(2)}\left( B_{ab} - M
    \pmatrix{0& \cr &-3\cr}\omega f_{44} \right)H_b^{(2)} \,.
\label{h2}
\end{eqnarray}
To complete the doublet-triplet splitting we should take \cite{fut2}
\begin{eqnarray}
  B_{ab} \,\simeq \,-M M_{ab}\,.
  \label{B}
\end{eqnarray}
Then from the equations (\ref{h2m}) and (\ref{h2m'}), we can see that
a pair of light Higgs doublet (scalar) actually survive down to the
low energy.
\begin{eqnarray}
  (\ref{h2}) &\!\! = \!\!& -M \bar{H}_a^{(2)}\left( M_{ab}
    +\pmatrix{0& \cr &-3\cr}\omega f_{44} \right)H_b^{(2)} \\[2mm]
  &\!\! = \!\!& -M \bar{H}^{(2)'}\pmatrix{\mu& 0 \cr 0 & \mu' \cr}
  H^{(2)'}\,.
\end{eqnarray}
Taking into account the uncertainty in the condition (\ref{B}), here
we introduce a free parameter $\delta m^2_3$ in the boundary condition
for $m_3^2\,$;
\begin{eqnarray}
  m^2_3(M_G) =  - M \mu + \delta m^2_3 \,,\qquad (\, |\delta m_3^2|
  \>\raisebox{0.4ex}{\mbox{$<$}}
  \hspace{-1.8ex}\raisebox{-0.8ex}{\mbox{$\sim$}}\> M^2_{SUSY})
  \label{m3}
\end{eqnarray}
which plays an important role in the following analyses\footnote{This
  parameter is necessary in order to cause the radiative symmetry
  breaking. }.
With the above boundary conditions (\ref{bcg}) $\sim$ (\ref{bcm}) and
(\ref{m3}) for MSSM couplings, we analyze the low-energy predictions
in each case (1), (2).  Although these conditions are universal and
very restrictive, we can find the parameter region where these
predictions are consistent with the low-energy experimental values.
When the threshold corrections are neglected, the analysis
procedure is as follows.
\\

\noindent$\bullet$ case (1) : ( large $\tan \beta$ model )
\setcounter{footnote}{0}

In this case we have five free parameters, $g\,,\,M_G\,,\,M\,,\,\mu$
and $\delta m^2_3$.$\;$ Since we don't deal with the threshold
corrections in this section, we can treat $M_{SUSY}$ and $\tan \beta$
just like as free parameters in the procedure.  At first
we input $g\,,\,M_G\,,\,M_{SUSY}$ and $\tan \beta$, and run the
dimensionless couplings down to $M_Z$ by using 2-loop
$\beta$-functions \cite{2beta}.  Then we can tune these four input
parameters to reproduce the low-energy values which are consistent
with the experimental data \cite{exp},
\begin{eqnarray}
  \alpha_1(M_Z) &\!\! = \!\!& 0.01689 \pm 0.00005 \,,\\
  \alpha_2(M_Z) &\!\! = \!\!& 0.03322 \pm 0.00025 \,,\\
  \alpha_3(M_Z) &\!\! = \!\!& 0.12 \pm 0.01 \,,\\
  m_b(M_Z) = 3.1 \pm 0.4 \;\mbox{GeV}\,,&& m_\tau(M_Z) = 1.75 \pm 0.01
  \>\mbox{GeV} \,,\\
  \quad M_Z &\!\! = \!\!& 91.187\;\mbox{GeV} \,.
\end{eqnarray}
The dimensionfull parameters are not included in these
$\beta$-functions and we neglect the threshold corrections,
therefore the values of $M\,,\,\mu$ and $\delta m^2_3$ give no effects
to this tuning.
Next with the input value of $M$ in addition to the above parameter
set, we calculate the dimensionfull parameters at $M$ and tune $M$ so
that $M_{SUSY}$ in this set may be equal to a value of the following
quantity;
\begin{equation}
  M_{SB} \equiv \frac{1}{4}\left( 2 m_{{\tilde Q}_3}^2(M) + m_{{\tilde
        u}_3}^2(M) + m_{{\tilde d}_3}^2(M) \right)\,.
\end{equation}
Similarly, this adjustment of $M$ is independent of the input values
of $\mu$ and $\delta m^2_3$, since none of the $\beta$-functions of
other couplings contain $\rho$ and $m_3^2$ parameters.
In the last step, we tune $\mu$ and $\delta m^2_3$ so that the
low-energy Higgs potential actually realize the value of $\tan \beta$
in this parameter set and fulfill the constraints for radiative
electroweak symmetry breaking,
\begin{eqnarray}
  \tan^2 \beta &\!\! = \!\!& \frac{m_1^2 + \rho^2 +
    \frac{1}{2}M_Z^2}{m_2^2 + \rho^2 + \frac{1}{2}M_Z^2} \,, \\
  \sin 2\beta &\!\! = \!\!& \frac{ -2 m_3^2 }{m_1^2 + m_2^2 + 2\rho^2}
  \,.
\end{eqnarray}
At this stage, we should incorporate the 1-loop corrections to
the minimization of the Higgs potential in order to improve a
very sensitive dependence of VEV on the renormalization
point $Q$ \cite{higgs}.\footnote{To treat this more precisely, we
  should make use of the improving effective potential method
  \cite{bando}.}
\ The 1-loop corrected Higgs potential is
\begin{eqnarray}
  V = V^{(0)} + V^{(1)}\,,
\end{eqnarray}
\vspace*{-7mm}
\begin{eqnarray}
  V^{(0)} &\!\! = \!\!& ( m_1^2 + \rho^2 ) \bar{H}^\dagger \bar{H} +
  ( m_2^2 + \rho^2 ) H^\dagger H + ( m_3^2 \bar{H} H + \mbox{ h.c.} )
  \\[2mm]
  && \qquad\quad  +\; ( \mbox{D-terms} ) \,, \nonumber \\[2mm]
  V^{(1)} &\!\! = \!\!& \frac{1}{64\pi^2} \,\mbox{STr}\left[ {\cal
      M}^4\left ( \mbox{ln} \frac{{\cal M}^2}{Q^2} - \frac{3}{2}
    \right)\right]\,,
\end{eqnarray}
where ${\cal M}^2$ is a field-dependent mass-squared matrix and
$\mbox{STr} \,{\cal A} = \sum_j \,(-1)^{2j}\,(2j+1) \,\mbox{Tr} {\cal
  A}_j$ is a weighted supertrace. The explicit expression for ${\cal
  M}^2$ can be found in Ref.$\,$\cite{higgsexp}.
This exact 1-loop correction takes, however, a very complicated form.
So we here adopt the handy calculating method \cite{handy} which
incorporates the corrections only to the Higgs mass terms from
$V^{(1)}$.  This method is known to be almost enough to obtain
approximate form of the full 1-loop potential, because
the values of VEVs evaluated by this method are almost the
same as those obtained from the full 1-loop potential and the rapid
$Q$ dependence of the potential which mainly comes from that of the
running mass parameters becomes milder even in this method
\cite{handy}.

In this way, we can determine the input values of
$g\,,\,M_G\,,\,M\,,\,\mu$ and $\delta m^2_3\,$, and calculate the
low-energy predictions (gauge couplings and masses of fermions,
superparticles and Higgs etc.). \
We show the two type of example in Table 2, 3 (One is the parameters
set for the highest $M_{SB}$ (Table 2) and another is lowest
one. (table 3)$\,$) where $m_{\tilde x}$ are the superparticle masses
and $m_{H^\pm}\,,\,m_A\,,\,m_{H, h}$ are the Higgs scalar
masses which correspond to the charged, neutral CP-odd and neutral
CP-even ones, respectively. Their explicit tree level forms are given
in the appendix A.

The sparticle mass predictions are enough within the experimental
bounds \cite{exp}.  Note that, unlike the usual MSSM, the lightest
superparticle(LSP) is not a neutral one but
$\tau$-slepton\footnote{This problem may be avoided by considering an
  $R$-parity violating interaction $Q \bar d L$ which is needed for
  one of the possible interpretations of the high-$Q^2$ anomaly at
  HERA \cite{hera}.} because of the highly restrictive finiteness
conditions and the largeness of $y_b$ and $y_\tau$.
This property is characteristic to this finite SU(5) model with large
$\tan \beta$ and will be tested in future experiments.

\vspace*{7mm}

\noindent$\bullet$ case (2) : ( small $\tan \beta$ model )

In a same way as case (1), we need five free parameters for
radiative symmetry breaking in this case. Unlike the large $\tan
\beta$ models, there is no problem of the fine-tuning of Higgs mass
parameters (and large threshold correction to $m_b$) in small $\tan
\beta$ models. On the other hand, if we adopt the handy calculating
method to estimate the 1-loop corrections to the Higgs potential, we
should take care of another respect as mentioned in
Ref.\cite{handy}. This is that in this case the contributions to
D-term from $V^{(1)}$ near the flat direction ($\tan
\beta \sim 1$) are no longer small compared with that from the tree
level $V^{(0)}$. Therefore this method which includes only mass
corrections may become no more good approximation.  The contribution
to the quartic terms from $V^{(0)}$ is
\begin{equation}
 \sim  \frac{1}{8}( g_1^2 + g_2^2 )|v|^4 \cos^2 2 \beta \,,
  \label{flat0}
\end{equation}
and the typical one from $V^{(1)}$ is
\begin{equation}
  \sim \frac{1}{32 \pi^2} g_2^2 |v|^4 \,.
 \label{flat1}
\end{equation}
Therefore requiring that for instance, (\ref{flat1}) is within
$10\,$\% of (\ref{flat0}), we need
\begin{equation}
  \tan \beta \;\raisebox{0.2ex}{\mbox{$>$}}
  \hspace{-1.8ex}\raisebox{-1.0ex}{\mbox{$\sim$}}\; 1.35 \,.
\end{equation}
As for all other respects, we follow the analysis procedure in case
(1) and the representative result is shown in Table 4.

We can see from this, for example, that this model predicts that the
LSP is the lightest neutralino because $\mu$ can be taken smaller than
that in the large $\tan \beta$ case etc.\ and therefore this result
bears a close resemblance to the typical one in usual MSSM
\cite{mssm}.

\newpage

\begin{center}
  \renewcommand{\arraystretch}{1.25} \setlength{\doublerulesep}{1mm}
  \begin{tabular}{|c|c||c|c|} \hline
    \multicolumn{4}{|c|}{\large case (1) : example 1} \\
    \hline\hline $M_{GUT}$ & \multicolumn{2}{c}{ $1.246\times 10^{16}$
      ~(GeV)} & \\ \hline $\alpha_{GUT}$ & \multicolumn{2}{c}{ 0.0388
      } & \\ \hline $M$ & \multicolumn{2}{c}{ 612.315 ~(GeV)} & \\
    \hline $\mu$ & \multicolumn{2}{c}{ 1467.828 ~(GeV)} & \\ \hline
    $\delta m_3^2$ & \multicolumn{2}{c}{ $- (636.086)^2$ ~(GeV)} & \\
    \hline\hline $M_{SB}$ & 1000.0 ~(GeV)& $\tan \beta$ & 54.0 \\
    \hline \makebox[2cm]{ $ \alpha_1(M_Z)$} & \makebox[4cm]{0.016850}
    & \makebox[2cm]{ $m_t(M_Z)$ } & \makebox[4cm]{ 179.24 ~(GeV)} \\
    \hline $\alpha_2(M_Z)$ & 0.033333 & $m_b(M_Z)$ & 3.21 ~(GeV)\\
    \hline $\alpha_3(M_Z)$ & 0.112 & $m_\tau(M_Z)$ & 1.745 ~(GeV)\\
    \hline \hline $m_{\widetilde{t}_+}$ & 1064.3 ~(GeV)&
    $m_{\widetilde{u}_+}$ & 1243.4 ~
    (GeV)\\ \hline $m_{\widetilde{t}_-}$ & 892.4 ~(GeV)&
    $m_{\widetilde{u}_-}$ &
    1198.3 ~(GeV)\\ \hline $m_{\widetilde{b}_+}$ & 1051.2 ~(GeV)&
    $m_{\widetilde{d}_+}$ & 1245.8 ~(GeV)\\ \hline
    $m_{\widetilde{b}_-}$ & 915.1 ~
    (GeV)& $m_{\widetilde{d}_-}$ & 1193.4 ~ (GeV)\\ \hline
    $m_{\widetilde{\tau}_+}$
    & 516.2 ~(GeV)& $m_{\widetilde{e}_+}$ & 536.2 ~(GeV)\\ \hline
    $m_{\widetilde{\tau}_-}$ & 191.2 ~(GeV)& $m_{\widetilde{e}_-}$ &
    420.5 ~ (GeV)\\ \hline $m_{\widetilde{\nu}_\tau}$ & 478.5 ~(GeV)&
    $m_{\widetilde{\nu}_e}$ & 530.6 ~(GeV)\\ \hline
    $m_{\widetilde{\chi}_1^+}$ &
    837.5 ~(GeV)& $m_{\widetilde{\chi}_2^+}$ & 481.5 ~(GeV)\\ \hline
    $m_{\widetilde{\chi}_1^0}$ & 275.4 ~ (GeV)&
    $m_{\widetilde{\chi}_2^0}$ & 500.5
    ~(GeV)\\ \hline $m_{\widetilde{\chi}_3^0}$ & 794.9 ~(GeV)&
    $m_{\widetilde{\chi}_4^0}$ & 814.9 ~(GeV)\\ \hline $m_{H^{\pm}}$ &
    331.8 ~(GeV)& $m_{A}$ & 322.6 ~(GeV)\\ \hline $m_{H}$ & 322.7 ~
    (GeV)& $m_{h}$ & 89.0 ~(GeV)\\ \hline $M_{\lambda_3}$ & 1326.6 ~
    (GeV)& & \\ \hline $h_t$ & $-$ 853.6 ~ (GeV)& $h_b$ & $-$ 834.0 ~
    (GeV)\\ \hline $h_\tau$ & $-$ 107.1 ~ (GeV)& & \\ \hline
\end{tabular}
\vspace*{1cm}

Table 2 : {The low-energy predictions in case (1). ~(high $M_{SB}$)}
\end{center}

\begin{center}
  \renewcommand{\arraystretch}{1.25} \setlength{\doublerulesep}{1mm}
  \begin{tabular}{|c|c||c|c|} \hline
    \multicolumn{4}{|c|}{\large case (1) : example 2} \\
    \hline\hline $M_{GUT}$ & \multicolumn{2}{c}{ $1.116\times 10^{16}$
      ~(GeV)} & \\ \hline $\alpha_{GUT}$ & \multicolumn{2}{c}{ 0.0392
      } & \\ \hline $M$ & \multicolumn{2}{c}{ 473.385 ~(GeV)} & \\
    \hline $\mu$ & \multicolumn{2}{c}{ 1171.954 ~(GeV)} & \\ \hline
    $\delta m_3^2$ & \multicolumn{2}{c}{ $- (510.579)^2$ ~(GeV)} & \\
    \hline\hline $M_{SB}$ & 790.0 ~(GeV)& $\tan \beta$ & 54.0 \\
    \hline \makebox[2cm]{ $ \alpha_1(M_Z)$} & \makebox[4cm]{0.016930}
    & \makebox[2cm]{ $m_t(M_Z)$ } & \makebox[4cm]{ 178.9 ~(GeV)} \\
    \hline $\alpha_2(M_Z)$ & 0.033462 & $m_b(M_Z)$ & 3.20 ~(GeV)\\
    \hline $\alpha_3(M_Z)$ & 0.113 & $m_\tau(M_Z)$ & 1.745 ~(GeV)\\
    \hline \hline $m_{\widetilde{t}_+}$ & 858.0 ~(GeV)&
    $m_{\widetilde{u}_+}$ & 978.3 ~
    (GeV)\\ \hline $m_{\widetilde{t}_-}$ & 694.5 ~(GeV)&
    $m_{\widetilde{u}_-}$ &
    944.2 ~(GeV)\\ \hline $m_{\widetilde{b}_+}$ & 841.5 ~(GeV)&
    $m_{\widetilde{d}_+}$ & 981.3 ~(GeV)\\ \hline
    $m_{\widetilde{b}_-}$ & 708.2 ~
    (GeV)& $m_{\widetilde{d}_-}$ & 940.9 ~ (GeV)\\ \hline
    $m_{\widetilde{\tau}_+}$
    & 415.0 ~(GeV)& $m_{\widetilde{e}_+}$ & 416.5 ~(GeV)\\ \hline
    $m_{\widetilde{\tau}_-}$ & 111.5 ~(GeV)& $m_{\widetilde{e}_-}$ &
    326.4 ~ (GeV)\\ \hline $m_{\widetilde{\nu}_\tau}$ & 369.4 ~(GeV)&
    $m_{\widetilde{\nu}_e}$ & 409.2 ~(GeV)\\ \hline
    $m_{\widetilde{\chi}_1^+}$ &
    666.4 ~(GeV)& $m_{\widetilde{\chi}_2^+}$ & 367.9 ~(GeV)\\ \hline
    $m_{\widetilde{\chi}_1^0}$ & 215.0 ~ (GeV)&
    $m_{\widetilde{\chi}_2^0}$ & 391.1
    ~(GeV)\\ \hline $m_{\widetilde{\chi}_3^0}$ & 613.3 ~(GeV)&
    $m_{\widetilde{\chi}_4^0}$ & 638.6 ~(GeV)\\ \hline $m_{H^{\pm}}$ &
    261.8 ~(GeV)& $m_{A}$ & 250.1 ~(GeV)\\ \hline $m_{H}$ & 250.2 ~
    (GeV)& $m_{h}$ & 88.9 ~(GeV)\\ \hline $M_{\lambda_3}$ & 1042.7 ~
    (GeV)& & \\ \hline $h_t$ & $-$ 677.2 ~ (GeV)& $h_b$ & $-$ 662.5 ~
    (GeV)\\ \hline $h_\tau$ & $-$ 78.5 ~ (GeV)& & \\ \hline
\end{tabular}
\vspace*{1cm}

Table 3 : {The low energy predictions in case (1). ~(low $M_{SB}$)}
\end{center}

\begin{center}
  \renewcommand{\arraystretch}{1.25} \setlength{\doublerulesep}{1mm}
  \begin{tabular}{|c|c||c|c|} \hline
    \multicolumn{4}{|c|}{\large case (2)} \\
    \hline\hline $M_{GUT}$ & \multicolumn{2}{c}{ $1.206\times 10^{16}$
      ~(GeV)} & \\ \hline $\alpha_{GUT}$ & \multicolumn{2}{c}{ 0.0392
      } & \\ \hline $M$ & \multicolumn{2}{c}{ 332.38 ~(GeV)} & \\
    \hline $\mu$ & \multicolumn{2}{c}{ 642.65 ~(GeV)} & \\ \hline
    $\delta m_3^2$ & \multicolumn{2}{c}{ $- (255.15)^2$ ~(GeV)} & \\
    \hline\hline $M_{SB}$ & 620.0 ~(GeV)& $\tan \beta$ & 3.1 \\ \hline
    \makebox[2cm]{ $ \alpha_1(M_Z)$} & \makebox[4cm]{0.016852} &
    \makebox[2cm]{ $m_t(M_Z)$ } & \makebox[4cm]{ 177.5 ~(GeV)} \\
    \hline $\alpha_2(M_Z)$ & 0.033119 & $m_b(M_Z)$ & 3.42 ~(GeV)\\
    \hline $\alpha_3(M_Z)$ & 0.110 & $m_\tau(M_Z)$ & 1.751 ~(GeV)\\
    \hline \hline $m_{\widetilde{t}_+}$ & 685.5 ~(GeV)&
    $m_{\widetilde{u}_+}$ & 690.8 ~
    (GeV)\\ \hline $m_{\widetilde{t}_-}$ & 453.8 ~(GeV)&
    $m_{\widetilde{u}_-}$ &
    667.4 ~(GeV)\\ \hline $m_{\widetilde{b}_+}$ & 665.8 ~(GeV)&
    $m_{\widetilde{d}_+}$ & 694.4 ~(GeV)\\ \hline
    $m_{\widetilde{b}_-}$ & 619.2 ~
    (GeV)& $m_{\widetilde{d}_-}$ & 665.8 ~ (GeV)\\ \hline
    $m_{\widetilde{\tau}_+}$
    & 293.6 ~(GeV)& $m_{\widetilde{e}_+}$ & 292.8 ~(GeV)\\ \hline
    $m_{\widetilde{\tau}_-}$ & 229.0 ~(GeV)& $m_{\widetilde{e}_-}$ &
    231.0 ~ (GeV)\\ \hline $m_{\widetilde{\nu}_\tau}$ & 284.4 ~(GeV)&
    $m_{\widetilde{\nu}_e}$ & 284.4 ~(GeV)\\ \hline
    $m_{\widetilde{\chi}_1^+}$ &
    559.0 ~(GeV)& $m_{\widetilde{\chi}_2^+}$ & 246.3 ~(GeV)\\ \hline
    $m_{\widetilde{\chi}_1^0}$ & 188.7 ~ (GeV)&
    $m_{\widetilde{\chi}_2^0}$ & 296.9
    ~(GeV)\\ \hline $m_{\widetilde{\chi}_3^0}$ & 411.5 ~(GeV)&
    $m_{\widetilde{\chi}_4^0}$ & 493.9 ~(GeV)\\ \hline $m_{H^{\pm}}$ &
    654.1 ~(GeV)& $m_{A}$ & 649.5 ~(GeV)\\ \hline $m_{H}$ & 651.6 ~
    (GeV)& $m_{h}$ & 72.0 ~(GeV)\\ \hline $M_{\lambda_3}$ & 734.5 ~
    (GeV)& & \\ \hline $h_t$ & $-$ 523.1 ~ (GeV)& $h_b$ & $-$ 63.2 ~
    (GeV)\\ \hline $h_\tau$ & $-$ 17.6 ~ (GeV)& & \\ \hline
\end{tabular}
\vspace*{1cm}

Table 4 : {The low energy predictions in case (2).}
\end{center}

\section{Threshold corrections}
\setcounter{equation}{0}\setcounter{footnote}{0}

In this section we address the threshold corrections at GUT and SUSY
scales. Two types of corrections are important$\,$; One is the GUT
corrections to gauge and Yukawa couplings \cite{gth,gsth} which may be
characteristic to this finite SU(5).\footnote{The GUT corrections to
  the other parameters (gaugino masses etc.) are so small that the
  following analysis is not affected too much \cite{gutth}.}
Another is the SUSY ones \cite{gsth,sth}, especially the corrections
to $m_b$ which are important in the large $\tan \beta$ model.
These corrections to the dimensionless couplings are thought to be
important in the sense that these parameters are now precisely
measured by experiments and so including these corrections may
restrict the models and their allowed parameter regions. However,
since the MSSM couplings are highly restricted at GUT scale by
considering the finiteness conditions and the low-energy physics, that
is, the experimental data and the constraints from the Higgs
potential, there is only a little room for varying the SUSY threshold
corrections except for these signs. Therefore we consider whether the
low-energy experimental values are consistently reproduced by tuning
the GUT threshold corrections. In this paper we neglect the 1-loop
corrections from electroweak gauge boson and top quark \cite{weak}
except for the important and large corrections to the mass of the
lightest CP-even Higgs boson \cite{higgsth}.

First, we discuss the 1-loop GUT threshold corrections.
These corrections to the standard gauge couplings are found to be
\begin{eqnarray}
  \frac{2\pi}{\alpha_i(\Lambda)} = \,\frac{2\pi}{\alpha_5}\, -
  \Delta_i^G(\Lambda)\,,
\end{eqnarray}
\begin{eqnarray}
  \Delta_1^G(\Lambda) &\!\! = \!\!&  \frac{5}{2} \ln \left
    ( \frac{M_V}{\Lambda} \right) -
  \frac{1}{4}\sum_{i = 1, 2} \ln \left( \frac{M_i}{\Lambda} \right)  -
  \frac{1}{10} \sum_{i = 3, 4} \ln \left( \frac{M_i}{\Lambda} \right)
  - \frac{3}{20} \ln \left( \frac{\mu'}{\Lambda} \right) ,\; \\[1mm]
  \Delta_2^G(\Lambda) &\!\! = \!\!&  \frac{3}{2} \ln \left
    ( \frac{M_V}{\Lambda} \right) -
  \frac{1}{2} \ln \left( \frac{M_\Sigma}{\Lambda} \right) -
  \frac{1}{4}\sum_{i = 1, 2} \ln \left( \frac{M_i}{\Lambda} \right) -
  \frac{1}{4} \ln \left( \frac{\mu'}{\Lambda} \right) \,,\\[1mm]
  \Delta_3^G(\Lambda) &\!\! = \!\!&  \ln \left( \frac{M_V}{\Lambda}
  \right) - \frac{4}{3} \ln \left( \frac{M_\Sigma}{\Lambda} \right) -
  \frac{1}{4}\sum_{i = 1, 2, 3, 4} \ln \left( \frac{M_i}{\Lambda}
  \right) \,,
\end{eqnarray}
where $M_V$ is the superheavy gauge bosons mass (which is
equivalent to that of the $(3,2,\pm\frac{5}{6})$ component of adjoint
Higgs $\Sigma\,$),~ $M_i$ and $\mu'$ are the masses of
superheavy parts of $H_a\,,\,{\bar H}_a$, and $M_\Sigma$ is the mass
of the color octet and SU(2) triplet component of $\Sigma$.
These mass parameters are defined by the conditions of SU(5)
symmetry breaking and the finiteness conditions as follows;
\begin{eqnarray}
  M_V &\!\! = \!\!&  5 \sqrt{2} g \omega = \frac{10}{3} \sqrt
  { \frac{14}{15} } M_{24}  \,,\\[1.5mm]
  M_\Sigma &\!\! = \!\!& 5 M_{24} \,,\\
  M_3 &\!\! = \!\!& 5 \omega f_{44} \sin \theta \sin {\bar \theta} =
  \frac{10}{3} \sqrt{ \frac{7}{15} } \sin \theta \sin {\bar \theta}\,
  M_{24} \,,\\
  M_4 &\!\! = \!\!& \mu' + 5 \omega f_{44} \cos \theta \cos {\bar
    \theta}  = \mu' + \frac{10}{3} \sqrt{ \frac{7}{15} } \cos \theta
  \cos {\bar \theta}\, M_{24} \,.
\end{eqnarray}
$M_1\,,\,M_2\,,\,\mu'$ and $\omega$ are defined in section 3
($\,$(\ref{m12}) and
(\ref{h2m'})$\,$). $\,M_{24}$ is the supersymmetric mass parameter of
$\Sigma$,
\begin{eqnarray}
  W_m = W'_m + M_{24} \,\Sigma^{\,2}.
\end{eqnarray}
Since we set $M_G = M_V = \Lambda$, there are three free parameters,
$M_1\,,\,M_2$ and $\mu'$ in the correction formulae. Furthermore since
$H_1$ and $H_2$ sector have same structures, hereafter without loss of
generality we set $M_1 = M_2 \equiv M_H$. For the example of case 1
(large $\tan \beta$) and case 2 (small $\tan \beta$), the allowed
region for $M_H$ and $\mu'$ which reproduce the experimental values of
the gauge couplings $\alpha_{1,2,3}$ are shown in Fig.1,$\,$2
including the SUSY threshold corrections to gauge couplings (appendix
B). These SUSY threshold corrections $\Delta^S_{1,2,3}$ are about $1
\sim 2$ \%. Therefore it is found that these can be cancelled by
$\Delta^G_{1,2,3}\,$. However
this region may be rather narrowed if the proton decay constraint
$(M_H\,,\,\mu'\>\raisebox{0.2ex}{\mbox{$>$}}
\hspace{-1.8ex}\raisebox{-1.0ex}{\mbox{$\sim$}}\> 10^{16}\,
\mbox{GeV})\,$ is considered.

\begin{figure}[htbp]
  \begin{center}
    \leavevmode
    \epsfxsize=10cm \epsfysize=8cm \ \epsfbox{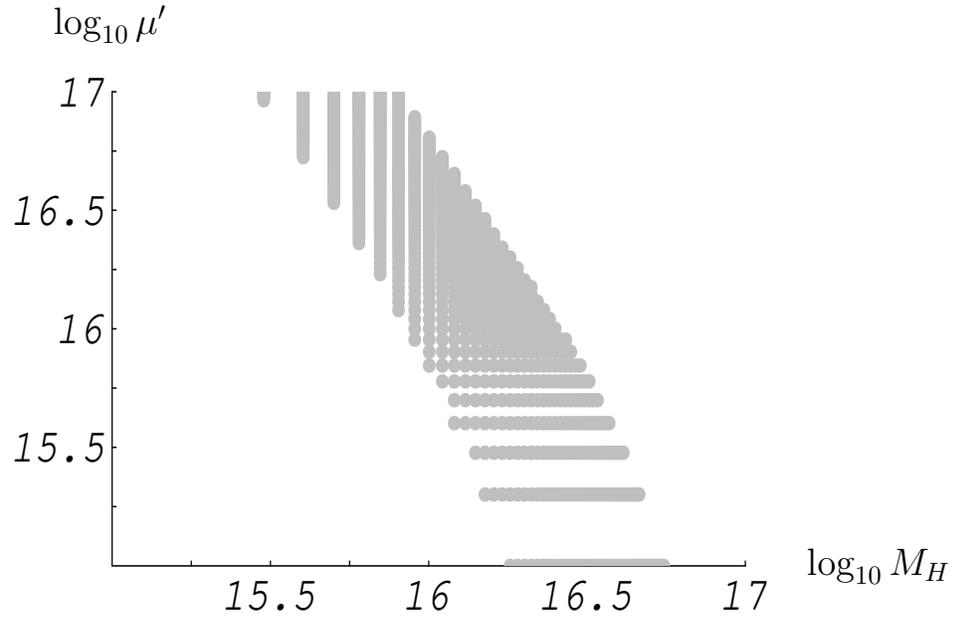}
    \put(15,25){\large $\log_{10} M_H$}
    \put(-270,230){\large $\log_{10} \mu'$}
    \caption{ The allowed region for $M_H$ vs. $\mu'$. (large $\tan
      \beta$ case)}
  \end{center}
\end{figure}

\begin{figure}[htbp]
  \begin{center}
    \leavevmode
    \epsfxsize=10cm \epsfysize=8cm \ \epsfbox{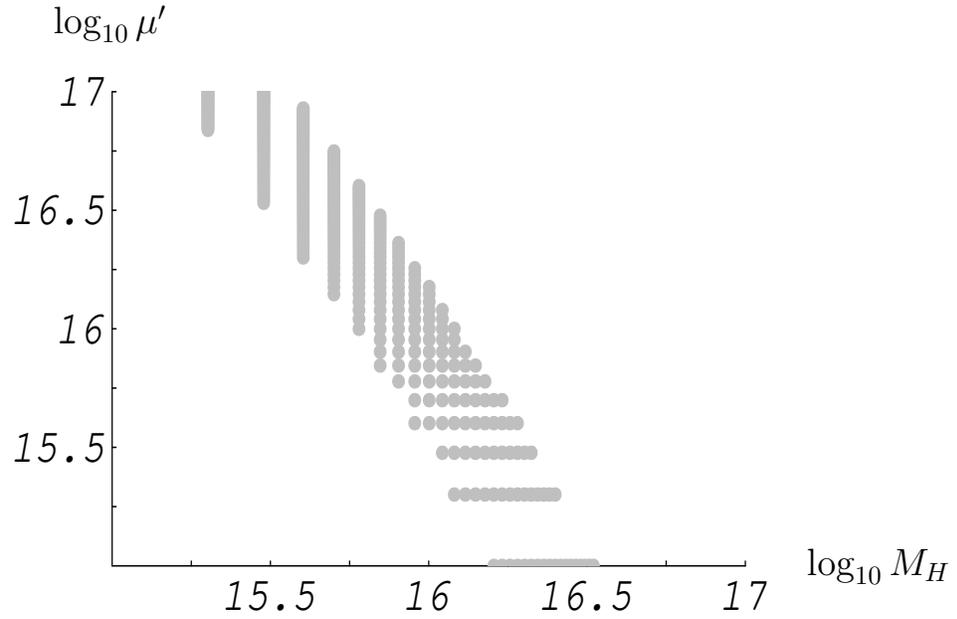}
    \put(15,25){\large $\log_{10} M_H$}
    \put(-270,230){\large $\log_{10} \mu'$}
    \caption{ The allowed region for $M_H$ vs. $\mu'$. (small $\tan
      \beta$ case)}
  \end{center}
\end{figure}

Next we consider the GUT corrections to Yukawa couplings,
\begin{eqnarray}
  y_t^-( \Lambda ) &\!\! = \!\!& y_t^+ ( 1 + \Delta_t^G (\Lambda))\,,
  \\
  y_b^-( \Lambda ) &\!\! = \!\!& y_b^+ ( 1 + \Delta_b^G (\Lambda))\,,
  \\
  y_\tau^-( \Lambda ) &\!\! = \!\!& y_\tau^+ ( 1 + \Delta_\tau^G
  (\Lambda))\,,
\end{eqnarray}
\begin{eqnarray}
  16 \pi^2 \Delta_t^G(\Lambda) &\!\! = \!\!& - \frac{g^2}{2} \left( 5
    F(M_V^2,0) + 3
    F(M_3^2,M_V^2) \right) + \frac{3}{2} \left( y_t^{+2} \cos^2 \theta
    + y_b^{+2} \cos^2 {\bar \theta}
  \right) F(M_3^2,0) \nonumber \\
  && + \frac{3}{2} \left( y_t^{+2} \sin^2 \theta +
    y_b^{+2} \sin^2 {\bar \theta} \right) F(M_4^2,0) \nonumber \\[1mm]
  && + \frac{1}{2} f_{44}^2 \sin^2 \theta \sin^2 {\bar \theta} \left
    ( 3 F(M_3^2,M_V^2) + \frac{3}{2} F(M_\Sigma^2,0) + \frac{3}{10}
    F((0.2M_\Sigma)^2,0) \right) \nonumber \\[1mm]
 && + \frac{1}{2} f_{44}^2 \left( \cos^2 \theta \sin^2 {\bar \theta} +
   \sin^2 \theta \cos^2 {\bar \theta} \right) \left( 3
    F(M_4^2,M_V^2) \right. \nonumber \\[1mm]
  && \left. + \frac{3}{2} F(M_\Sigma^2,\mu'^2) + \frac{3}{10}
    F((0.2M_\Sigma)^2,\mu'^2) \right) \,,
\end{eqnarray}
\begin{eqnarray}
  16 \pi^2 \Delta_b^G(\Lambda) &\!\! = \!\!& - \frac{g^2}{2} \left( 5
    F(M_V^2,0) + 3 F(M_3^2,M_V^2) \right) +  \left( y_t^{+2} \cos^2
    \theta + \frac{3}{2} y_b^{+2} \cos^2 {\bar \theta}
  \right) F(M_3^2,0) \nonumber \\
  && + \left( y_t^{+2} \sin^2 \theta + \frac{3}{2}
    y_b^{+2} \sin^2 {\bar \theta} \right) F(M_4^2,0) \nonumber \\[1mm]
  && + \frac{1}{2} f_{44}^2 \sin^2 \theta \sin^2 {\bar \theta} \left
    ( 3 F(M_3^2,M_V^2) + \frac{3}{2} F(M_\Sigma^2,0) + \frac{3}{10}
    F((0.2M_\Sigma)^2,0) \right) \nonumber \\[1mm]
  && + \frac{1}{2} f_{44}^2 \left( \cos^2 \theta \sin^2 {\bar \theta} +
    \sin^2 \theta \cos^2 {\bar \theta} \right) \left( 3 F(M_4^2,M_V^2)
  \right. \nonumber \\[1mm]
  && \left. + \frac{3}{2} F(M_\Sigma^2,\mu'^2) + \frac{3}{10}
    F((0.2M_\Sigma)^2,\mu'^2) \right) \,,
\end{eqnarray}
\begin{eqnarray}
  16 \pi^2 \Delta_\tau^G(\Lambda) &\!\! = \!\!& - \frac{g^2}{2} \left
    ( 9 F(M_V^2,0) + 3 F(M_3^2,M_V^2) \right) +  \frac{3}{2} \left
    ( y_t^{+2} \cos^2 \theta + y_b^{+2} \cos^2 {\bar \theta}
  \right) F(M_3^2,0) \nonumber \\
  && +  \frac{3}{2} \left( y_t^{+2} \sin^2 \theta +
    y_b^{+2} \sin^2 {\bar \theta} \right) F(M_4^2,0) \nonumber \\[1mm]
  && + \frac{1}{2} f_{44}^2 \sin^2 \theta \sin^2 {\bar \theta} \left
    ( 3 F(M_3^2,M_V^2) + \frac{3}{2} F(M_\Sigma^2,0) + \frac{3}{10}
    F((0.2M_\Sigma)^2,0) \right) \nonumber \\[1mm]
 && + \frac{1}{2} f_{44}^2 \left( \cos^2 \theta \sin^2 {\bar \theta} +
   \sin^2 \theta \cos^2 {\bar \theta} \right) \left( 3
    F(M_4^2,M_V^2) \right. \nonumber \\[1mm]
  && \left. + \frac{3}{2} F(M_\Sigma^2,\mu'^2) + \frac{3}{10}
    F((0.2M_\Sigma)^2,\mu'^2) \right)
\end{eqnarray}
where the superscript $+$ denotes that these couplings are GUT scale
parameters. From the finiteness conditions (\ref{sol1}) we have,
\begin{eqnarray}
  y_t^+ = \sqrt{\frac{8}{5}}\; g\,,\qquad y_b^+ = \sqrt{\frac{6}{5}}\;
  g\,,\qquad f_{44} = g \,.
\end{eqnarray}
The threshold function $F(a,b)$ is defined as follows;
\begin{eqnarray}
  F(m^2_a,m^2_b) &\!\! = \!\!& \frac{1}{ m^2_a - m^2_b } \left( m^2_a
    \ln \left( \frac{m^2_a}{\Lambda^2} \right) - m^2_b \ln \left(
      \frac{m^2_b}{\Lambda^2} \right) \right) - 1 \,.
\end{eqnarray}
The typical values of these corrections to the low-energy fermion
masses and the bottom-tau ratio are shown in Fig.3,$\,$5.

\begin{figure}[htbp]
  \parbox{7cm}{
    \epsfxsize=7cm \epsfysize=7cm \ \epsfbox{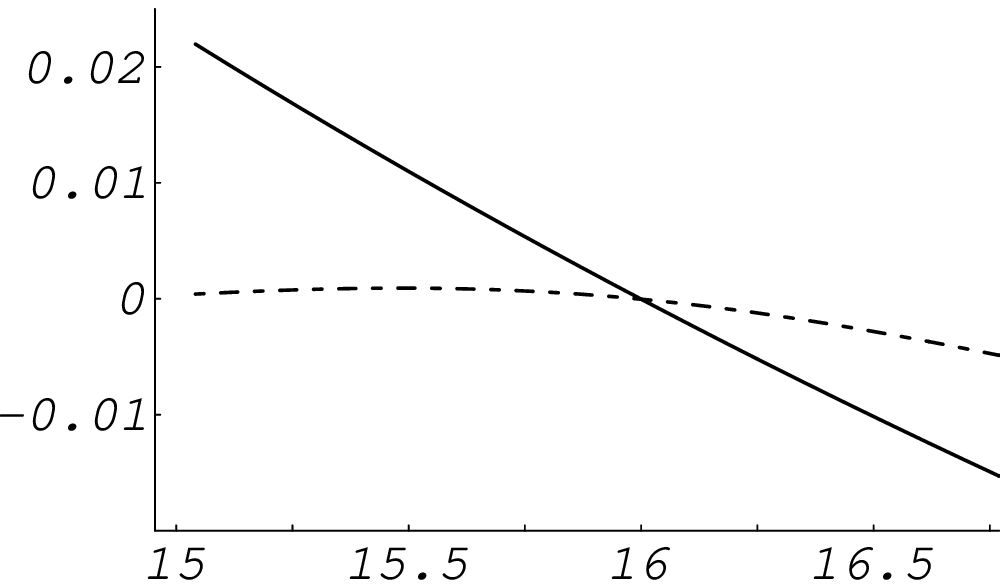}
    \put(-190,210){$\displaystyle{\frac{\Delta m_t(M_Z)}{m_t(M_Z)}}$}
    }
  \hspace*{8mm}
  \parbox{7cm}{
    \epsfxsize=7cm \epsfysize=7cm \ \epsfbox{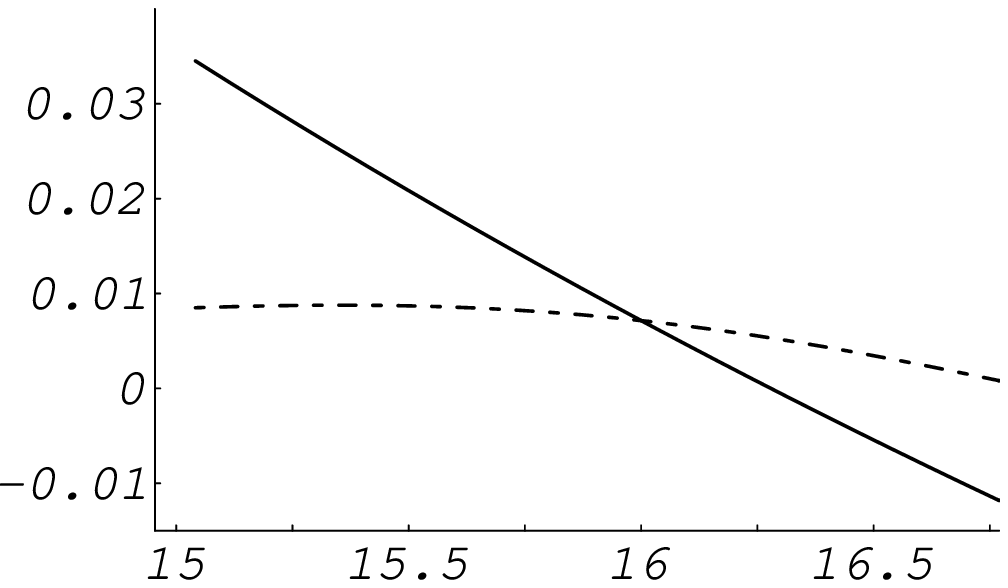}
    \put(-190,210){$\displaystyle{\frac{\Delta m_b(M_Z)}{m_b(M_Z)}}$}
    }
\end{figure}

\begin{figure}[htbp]
  \vspace*{-10mm}
  \parbox{7cm}{
    \epsfxsize=7cm \epsfysize=7cm \ \epsfbox{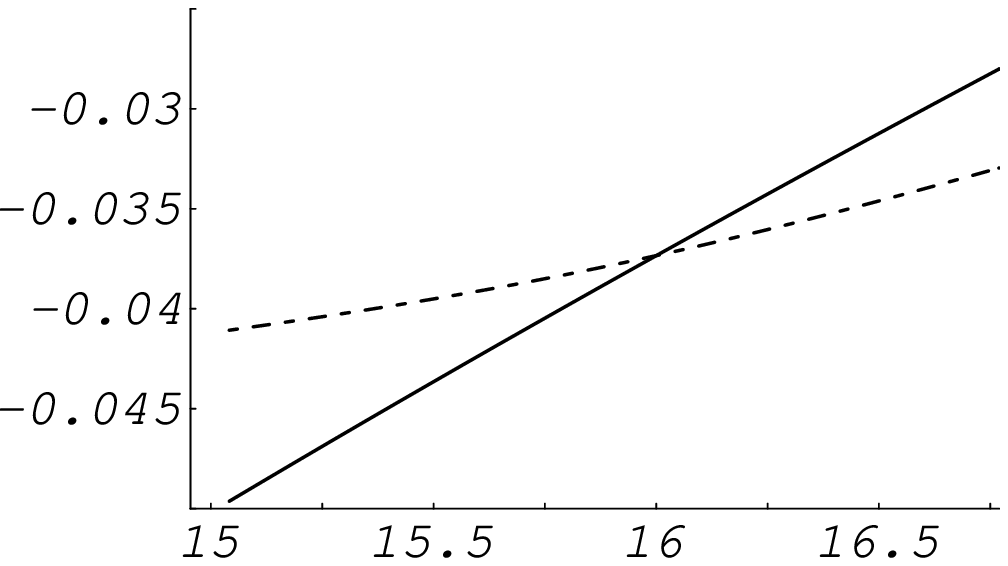}
    \put(-185,205){$\displaystyle{\frac{\Delta
          m_\tau(M_Z)}{m_\tau(M_Z)}}$}
    }
  \hspace*{8mm}
  \parbox{7cm}{
    \epsfxsize=7cm \epsfysize=7cm \ \epsfbox{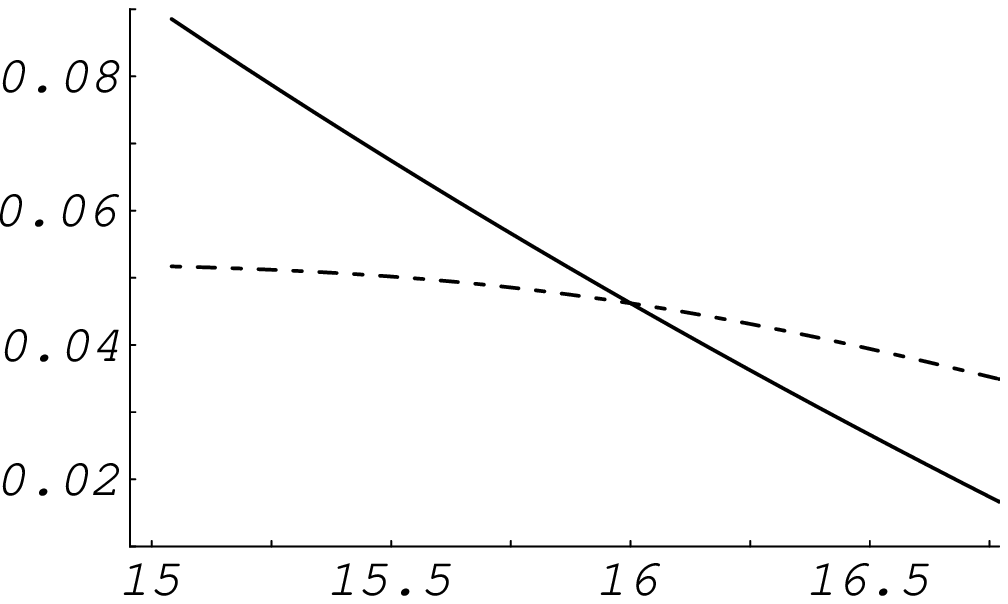}
    \put(-200,210){$\displaystyle{\frac{\Delta
          R_{b/\tau}(M_Z)}{R_{b/\tau}(M_Z)}}$}
    }
  \vspace*{3mm}
  \caption{The GUT threshold corrections to fermion masses. (large
  $\tan \beta$ case) ~The \hspace*{1.7cm}solid and dashed lines
  indicate $M_H$ and $\mu'$ dependences.}
\end{figure}

\begin{figure}[htbp]
  \begin{center}
    \leavevmode
    \epsfxsize=10cm \epsfysize=8cm \ \epsfbox{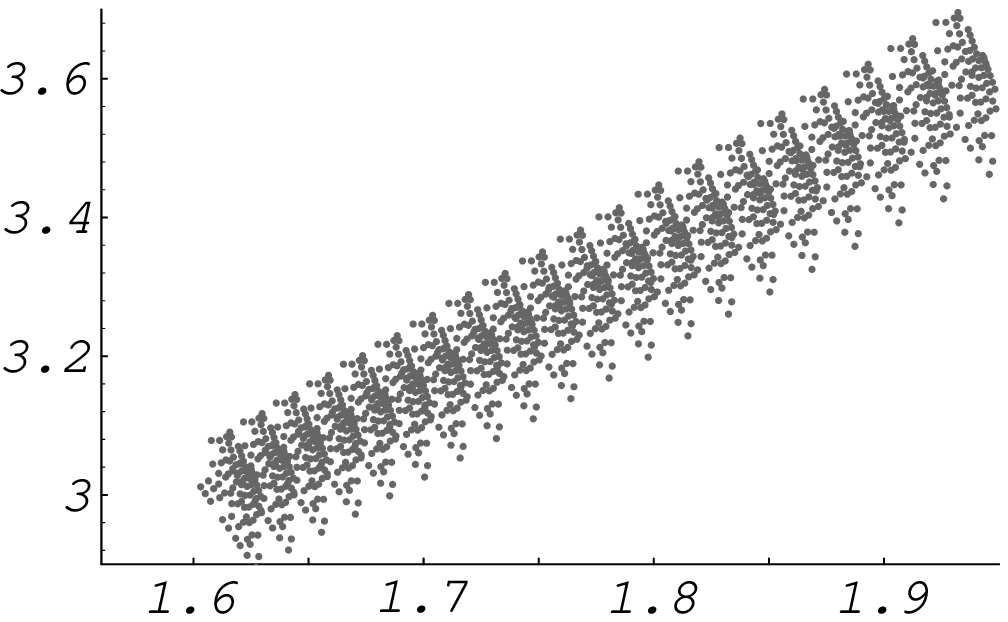}
    \put(15,25){\large $m_\tau(M_Z)$}
    \put(-270,230){\large $m_b(M_Z)$}
 \end{center}
 \vspace*{-5mm}
 \caption{ The typical values of $m_b(M_Z)$ and $m_\tau(M_Z)$ for the
 allowed parameter space in large $\tan \beta$ case.}
\end{figure}

We first investigate the large $\tan \beta$ case. In this case, the
SUSY threshold correction to $m_b(M_Z)$ is very large (about $25\,$\%)
for the contribution from gluino/squark and chargino/squark diagrams
due to large $\alpha_3$ and $y_t$, especially in models with the
universal soft SUSY breaking terms like this model \cite{bot,sth}. The
sign of this important correction, however, depends on that of the
supersymmetric Higgs mass parameter $\mu$ which can be easily
changed. This change gives only a slight effects to the low-energy
parameters and the SUSY threshold corrections. The SUSY correction to
$m_\tau(M_Z)$ is also important and about $5\,$\% which always has the
opposite sign to that to $m_b$ in large $\tan \beta$ case. Taking into
account the above facts, though $m_b(M_Z)$ has an experimental
uncertainty of about $15\,$\%, it is found from Fig.3 and 4 that in
this case we cannot predict the proper value of $m_b(M_Z)$.

On the other hand in small $\tan \beta$ case, the SUSY threshold
corrections $\Delta^S_{t,\,b, \tau}$ are all about a few
percents. Therefore we may tune $\Delta^G(M_H\,,\,\mu')$ so that
the low-energy values may be properly reproduced. However in this
case, since $y_{ b, \,\tau}$ are small the ratio of bottom and tau
masses, $R_{b/\tau}$, becomes large and to make matters worse the
correction $\Delta^S_b$ always makes a positive contribution which is
independent of the sign of $\mu$-parameter. Then the experimental
bound $R_{b/\tau}(M_Z) \>\raisebox{0.4ex}{\mbox{$<$}}
\hspace{-1.8ex}\raisebox{-0.8ex}{\mbox{$\sim$}}\> 2.0$ highly
constrains the parameter spaces. In Fig.2, by considering these
experimental constraints, only the left and above region of the gray
one is allowed. An example in this allowed parameter space is
shown in Table 5 in which the 1-loop corrections to the mass of the
lightest Higgs boson, $m_h$, are also included. However it seems that
only very narrow parameter regions are left even in this case due to
the correction to $m_b$ and $m_\tau$.

\begin{figure}[htbp]
  \parbox{7cm}{
    \epsfxsize=7cm \epsfysize=7cm \ \epsfbox{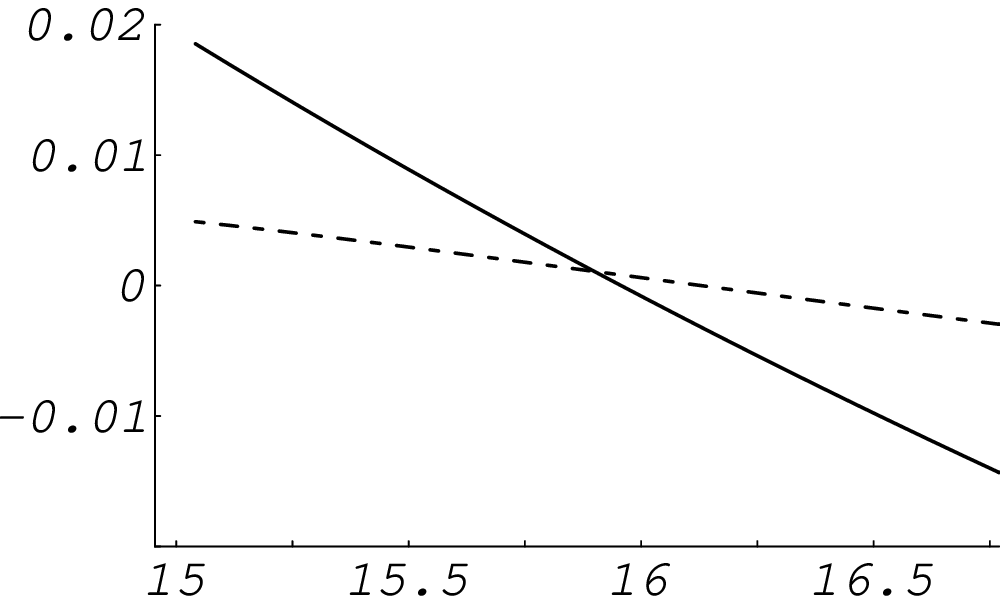}
    \put(-190,210){$\displaystyle{\frac{\Delta m_t(M_Z)}{m_t(M_Z)}}$}
    }
  \hspace*{8mm}
  \parbox{7cm}{
    \epsfxsize=7cm \epsfysize=7cm \ \epsfbox{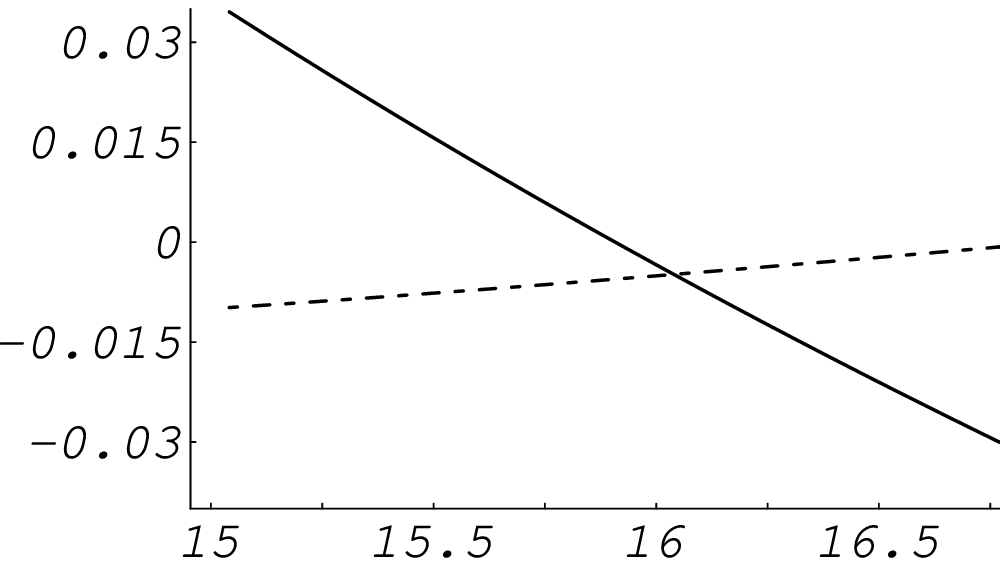}
    \put(-190,210){$\displaystyle{\frac{\Delta m_b(M_Z)}{m_b(M_Z)}}$}
    }
\end{figure}

\begin{figure}[htbp]
  \vspace*{-10mm}
  \parbox{7cm}{
    \epsfxsize=7cm \epsfysize=7cm \ \epsfbox{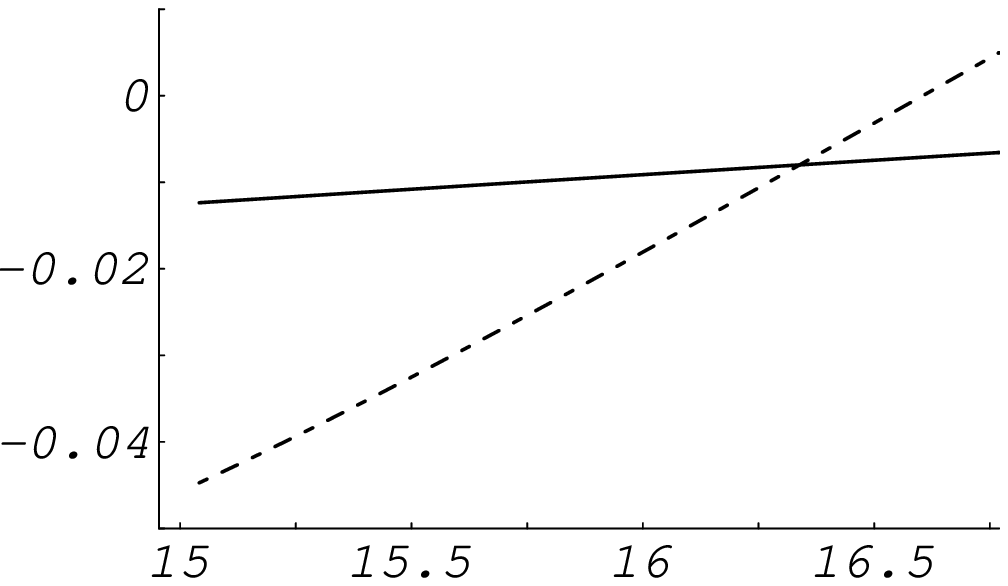}
    \put(-185,205){$\displaystyle{\frac{\Delta
          m_\tau(M_Z)}{m_\tau(M_Z)}}$}
    }
  \hspace*{8mm}
  \parbox{7cm}{
    \epsfxsize=7cm \epsfysize=7cm \ \epsfbox{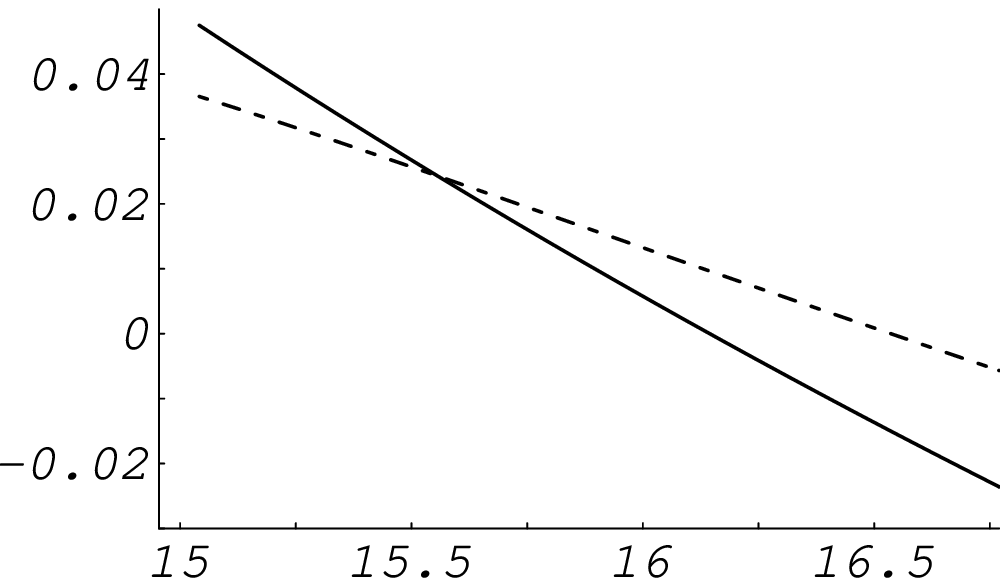}
    \put(-200,210){$\displaystyle{\frac{\Delta
          R_{b/\tau}(M_Z)}{R_{b/\tau}(M_Z)}}$}
  }
  \vspace*{3mm}
  \caption{The GUT threshold corrections to fermion masses. (small
  $\tan \beta$ case) ~The \hspace*{1.7cm}solid and dashed lines
  indicate $M_H$ and $\mu'$ dependences.}
\end{figure}

\begin{center}
  \renewcommand{\arraystretch}{1.25} \setlength{\doublerulesep}{1mm}
  \begin{tabular}{|c|c||c|c|} \hline
    \multicolumn{4}{|c|}{\large case (2) with threshold corrections} \\
    \hline\hline $M_{GUT}$ & \multicolumn{2}{c}{ $1.206\times 10^{16}$
      ~(GeV)} & \\ \hline $\alpha_{GUT}$ & \multicolumn{2}{c}{ 0.0392
      } & \\ \hline $M$ & \multicolumn{2}{c}{ 332.38 ~(GeV)} & \\
    \hline $\mu$ & \multicolumn{2}{c}{ 642.65 ~(GeV)} & \\ \hline
    $\delta m_3^2$ & \multicolumn{2}{c}{ $- (255.15)^2$ ~(GeV)} & \\
    \hline\hline $M_{SB}$ & 620.0 ~(GeV)& $\tan \beta$ & 3.1 \\ \hline
    \makebox[2cm]{ $ \alpha_1(M_Z)$} & \makebox[4cm]{0.016888} &
    \makebox[2cm]{ $m_t(M_Z)$ } & \makebox[4cm]{ 180.5 ~(GeV)} \\
    \hline $\alpha_2(M_Z)$ & 0.033014 & $m_b(M_Z)$ & 3.49 ~(GeV)\\
    \hline $\alpha_3(M_Z)$ & 0.115 & $m_\tau(M_Z)$ & 1.747 ~(GeV)\\
    \hline \hline $m_{\widetilde{t}_+}$ & 694.7 ~(GeV)&
    $m_{\widetilde{u}_+}$ & 700.2 ~
    (GeV)\\ \hline $m_{\widetilde{t}_-}$ & 461.5 ~(GeV)&
    $m_{\widetilde{u}_-}$ &
    677.2 ~(GeV)\\ \hline $m_{\widetilde{b}_+}$ & 675.6 ~(GeV)&
    $m_{\widetilde{d}_+}$ & 703.7 ~(GeV)\\ \hline
    $m_{\widetilde{b}_-}$ & 627.8 ~
    (GeV)& $m_{\widetilde{d}_-}$ & 675.6 ~ (GeV)\\ \hline
    $m_{\widetilde{\tau}_+}$
    & 293.7 ~(GeV)& $m_{\widetilde{e}_+}$ & 292.9 ~(GeV)\\ \hline
    $m_{\widetilde{\tau}_-}$ & 229.0 ~(GeV)& $m_{\widetilde{e}_-}$ &
    231.1 ~ (GeV)\\ \hline $m_{\widetilde{\nu}_\tau}$ & 284.5 ~(GeV)&
    $m_{\widetilde{\nu}_e}$ & 284.5 ~(GeV)\\ \hline
    $m_{\widetilde{\chi}_1^+}$ &
    560.8 ~(GeV)& $m_{\widetilde{\chi}_2^+}$ & 245.9 ~(GeV)\\ \hline
    $m_{\widetilde{\chi}_1^0}$ & 188.2 ~ (GeV)&
    $m_{\widetilde{\chi}_2^0}$ & 296.4
    ~(GeV)\\ \hline $m_{\widetilde{\chi}_3^0}$ & 411.8 ~(GeV)&
    $m_{\widetilde{\chi}_4^0}$ & 495.1 ~(GeV)\\ \hline $m_{H^{\pm}}$ &
    650.7 ~(GeV)& $m_{A}$ & 646.1 ~(GeV)\\ \hline $m_{H}$ & 648.2 ~
    (GeV)& $m_{h}$ & 94.8 ~(GeV)\\ \hline $M_{\lambda_3}$ & 743.7 ~
    (GeV)& & \\ \hline $h_t$ & $-$ 537.3 ~ (GeV)& $h_b$ & $-$ 64.9 ~
    (GeV)\\ \hline $h_\tau$ & $-$ 17.6 ~ (GeV)& & \\ \hline
\end{tabular}
\vspace*{5mm}

\begin{list}{}{}
\item \hspace*{-7mm} Table 5 : The low energy predictions in case (2)
  including GUT and SUSY threshold corrections. ($M_H = 0.4 \times
  10^{16}$ GeV and $\mu' = 2.5 \times 10^{16}$ GeV)
\end{list}
\end{center}

\section{Summary and Comments}
\setcounter{equation}{0}\setcounter{footnote}{0}

We discussed the SU(5) model with the finiteness conditions. In
this model $\beta$-functions of gauge and Yukawa couplings (and
supersymmetric mass parameters) are zero in all-order of perturbation
theory and the all other $\beta$-functions are zero in, at least,
two-loop order. Especially we analyzed the low energy taking into
account the problem in the Higgs potential. That is, we checked
whether the Higgs potential actually fulfill the both constraints of
large $\tan \beta$ and radiative electroweak symmetry breaking
including the 1-loop corrections to the Higgs mass parameters from
heavy $(\sim M_{SUSY})$ sector. As a result, it is found that without
breaking the finiteness conditions a free parameter is needed to
satisfy these constraints.  We also estimated the GUT and SUSY
threshold corrections to the dimensionless couplings in this
model. Including these corrections left us very small
available parameter spaces in this model. \ In this paper, we
discussed the particular form of the Higgs mass matrix, and yet it is
interesting problem to analyze a more general form of this matrix in
order to investigate the proton decay constraint, light fermion
masses, the CP-violation and so on.
The alternative way to construct the realistic and restricted (GUT)
model is the coupling constant reduction method \cite{ccrgut} based on
renormalization-group invariant relations among couplings which are
the solutions of the so-called reduction equations \cite{ccr}. Though
with these relations the models is not necessarily finite, one can
reduce the number of free parameters in models and increase the
predictive power as well as models with finiteness
conditions. Moreover the application of this method to the soft SUSY
breaking sector in ordinary minimal SU(5) model leads non-universal
boundary conditions for soft mass parameters at GUT scale. Therefore
one may improve the problem of the too large value of the SUSY
threshold correction to $m_b$ in large $\tan \beta$ model unlike the
finite SU(5) model.  In either way, the success or failure of the
models and the determination of the allowed parameter region entirely
depend on the near future experiments.

\section*{Acknowledgments}

We wish to express our deep gratitude to M.Bando for many helpful
discussions and encouragements in writing this paper. We would like to
thank to J.Sato for useful comments on the manuscript. We are also
grateful to T.Kugo, N.Maekawa and K.Shizuya for useful comments on the
analysis.

\vspace*{5mm}
\noindent
{\bf Note added:} After completing this paper, we became aware of the
papers by D.R.T.\ Jones et al.\ in Refs.\ \cite{fut,2beta}.\ \
We would like to thank D.R.T.\ Jones for informing us of their works.

\vspace*{3cm}

\renewcommand{\thesection}{\Alph{section}}
\setcounter{section}{1}
\setcounter{equation}{0}\setcounter{footnote}{0}

{\Large\bf \hspace*{-7mm}Appendix A}{\large\sl \hspace*{1ex} tree
  level mass formulae in MSSM}
\\

In this appendix we express the tree level mass formulae for MSSM
particles \cite{mssm}. In the following formulae, we neglect the
Yukawa couplings for the first and second generations.
\\

\noindent $\bullet$ sfermion masses for the third generation
\begin{eqnarray}
  M_{\tilde{t}_\pm}^2 &\!\! = \!\!&  \frac{1}{2} \left( m^2_{\tilde
      Q_3} + m^2_{\tilde t} + 2 m_t^2 + \frac{1}{2} M_Z^2 \cos 2 \beta
  \right. \\[1mm] && \left. \pm \sqrt{ \left ( m^2_{\tilde Q_3} -
        m^2_{\tilde t} + \left( \frac{1}{2} - \frac{4}{3} \sin^2
          \theta_W \right) M_Z^2 \cos 2 \beta \right)^2 + 4 |m_t|^2
      \left| A_t - \rho \cot \beta \right|^2 } \,\right)\,, \nonumber
  \\[1mm]
  M_{\tilde{b}_\pm}^2 &\!\! = \!\!&  \frac{1}{2} \left( m^2_{\tilde
      Q_3} + m^2_{\tilde b} + 2 m_b^2 - \frac{1}{2} M_Z^2 \cos 2 \beta
  \right. \\[1mm] && \left. \pm \sqrt{ \left( m^2_{\tilde Q_3} -
        m^2_{\tilde b} - \left( \frac{1}{2} - \frac{2}{3} \sin^2
          \theta_W \right) M_Z^2 \cos 2 \beta \right)^2 + 4 |m_b|^2
      \left| A_b - \rho \tan \beta \right|^2 } \,\right)\,, \nonumber
  \\[1mm]
  M_{\tilde{\tau}_\pm}^2 &\!\! = \!\!&  \frac{1}{2} \left( m^2_{\tilde
      L_3} + m^2_{\tilde \tau} + 2 m_\tau^2 - \frac{1}{2} M_Z^2 \cos 2
    \beta \right. \\[1mm] && \left. \pm \sqrt{ \left( m^2_{\tilde L_3}
        - m^2_{\tilde \tau} - \left( \frac{1}{2} - 2 \sin^2 \theta_W
        \right) M_Z^2 \cos 2 \beta \right)^2 + 4 |m_\tau|^2
      \left| A_\tau - \rho \tan \beta \right|^2 }
    \,\right)\,. \nonumber
\end{eqnarray}

\noindent $\bullet$ sfermion masses for the first and second
generations
\begin{eqnarray}
  M_{\tilde{u}_+}^2 &\!\! = \!\!&  m^2_{\tilde Q_{1, 2}} + \left
    ( \frac{1}{2} - \frac{2}{3} \sin^2 \theta_W \right) M_Z^2 \cos 2
  \beta \,,\\[1mm]
  M_{\tilde{u}_-}^2 &\!\! = \!\!&  m^2_{\tilde u_{1, 2}} + \frac{2}{3}
  \sin^2 \theta_W M_Z^2 \cos 2 \beta \,,\\[1mm]
  M_{\tilde{d}_+}^2 &\!\! = \!\!&  m^2_{\tilde Q_{1, 2}} - \left
    ( \frac{1}{2} - \frac{1}{3} \sin^2 \theta_W \right) M_Z^2 \cos 2
  \beta \,,\\[1mm]
  M_{\tilde{d}_-}^2 &\!\! = \!\!&  m^2_{\tilde d_{1, 2}} - \frac{1}{3}
  \sin^2 \theta_W  M_Z^2 \cos 2 \beta \,,\\[1mm]
  M_{\tilde{e}_+}^2 &\!\! = \!\!&  m^2_{\tilde L_{1, 2}} - \left
    ( \frac{1}{2} - \sin^2 \theta_W \right) M_Z^2 \cos 2 \beta
  \,,\\[1mm]
  M_{\tilde{e}_-}^2 &\!\! = \!\!&  m^2_{\tilde e_{1, 2}} - \sin^2
  \theta_W M_Z^2 \cos 2 \beta \,.
\end{eqnarray}

\noindent $\bullet$ chargino masses
\begin{eqnarray}
  m^2_{\tilde{\chi}_{1, 2}} &\!\! = \!\!& \frac{1}{2} \left
    ( (M_{\lambda_2}^2+\rho^2+2M_W^2) \pm
    \sqrt{(M_{\lambda_2}^2+\rho^2+2M_W^2)^2-4(M_{\lambda_2}
      \rho-M_W^2\sin 2\beta)^2} \right) \,.  \nonumber \\[1mm]
  \
\end{eqnarray}

\noindent $\bullet$ neutralino masses

The neutralino mass terms are :
\begin{eqnarray}
  {\cal L}_{\mbox{nm}} \,=\, - \frac{1}{2} \pmatrix{\widetilde{B}_L &
    \widetilde{W}^3_L & \widetilde{H}^0_{1L} & \widetilde{H}^0_{2L}
    \cr} M_n \pmatrix{\widetilde{B}_L \cr \widetilde{W}^3_L \cr
    \widetilde{H}^0_{1L} \cr \widetilde{H}^0_{2L} \cr} +
  \mbox{h.c.}\,,
\end{eqnarray}
where
\begin{eqnarray}
  \hspace*{-7mm}M_n = \pmatrix{M_{\lambda_1} & 0 & -M_Z \sin \theta_W
    \cos \beta & M_Z \sin \theta_W \sin \beta \cr 0 & M_{\lambda_2} &
    M_Z \cos \theta_W \cos \beta & - M_Z \cos \theta_W \sin \beta \cr
    -M_Z \sin \theta_W \cos \beta & M_Z \cos \theta_W \cos \beta & 0 &
    - \rho \cr M_Z \sin \theta_W \sin \beta & - M_Z \cos \theta_W \sin
    \beta & -\rho & 0 \cr }~\,.
\end{eqnarray}
\\

For $M_{SUSY} \gg M_Z\,$, $\,$the eigenvalues of $M_n$ are given by
\begin{eqnarray}
  m_{\widetilde{\chi}^0_1} &\!\! = \!\!& M_{\lambda_1} + \frac{M_Z^2
    \sin^2 \theta_W}{M_{\lambda_1}^2 - \rho^2} \left( M_{\lambda_1} +
    \rho \sin 2\beta \right) \,,\\[1mm]
  m_{\widetilde{\chi}^0_2} &\!\! = \!\!& M_{\lambda_2} + \frac{M_Z^2
    \cos^2 \theta_W}{M_{\lambda2}^2 - \rho^2} \left( M_{\lambda_2} +
    \rho \sin 2\beta \right)\,,\\[1mm]
  m_{\widetilde{\chi}^0_3} &\!\! = \!\!& \rho + \frac{M_Z^2 ( 1+ \sin
    2\beta ) }{ 2 ( \rho - M_{\lambda_1} ) ( \rho - M_{\lambda_2} ) }
  \left( \rho - M_{\lambda_1} \cos^2 \theta_W -
    M_{\lambda_2} \sin^2 \theta_W \right) \,,\\[1mm]
  m_{\widetilde{\chi}^0_4} &\!\! = \!\!& \rho + \frac{M_Z^2 ( 1- \sin
    2\beta ) }{ 2 ( \rho + M_{\lambda_1} ) ( \rho + M_{\lambda_2} ) }
  \left( \rho + M_{\lambda_1} \cos^2 \theta_W + M_{\lambda_2}
    \sin^2 \theta_W \right) \,.
\end{eqnarray}
\\

\noindent $\bullet$ Higgs scalar masses
\begin{eqnarray}
  m^2_{H^\pm} \!&\!\! = \!\!& m_1^2 + m_2^2 + 2 \rho^2 + M_W^2
  \,,\\[2mm]
  m_{A}^2 \,&\!\! = \!\!& m_1^2 + m_2^2 + 2\rho^2 \,,\\[2mm]
  m^2_{H} \,&\!\! = \!\!& \frac{1}{2} \left( m^2_{A} + M_Z^2 + \sqrt
    { \left ( m^2_{A} + M_Z^2 \right)^2 - 4 m^2_{A} M_Z^2 \cos^2
      2\beta} \right) \,,\\[2mm]
  m^2_{h} \,&\!\! = \!\!& \frac{1}{2} \left( m^2_{A} + M_Z^2 - \sqrt
    { \left( m^2_{A} + M_Z^2 \right)^2 - 4 m^2_{A} M_Z^2 \cos^2
      2\beta} \right) \,.
\end{eqnarray}

\vspace*{1.5cm}

\setcounter{section}{2}
\setcounter{equation}{0}\setcounter{footnote}{0}

{\Large\bf \hspace*{-7mm}Appendix B}{\large\sl \hspace*{1ex} SUSY
  threshold corrections to gauge couplings}
\\

In this appendix we present the explicit forms of the SUSY threshold
corrections to the MSSM gauge couplings \cite{gsth,sth} in the same
approximation as appendix A. For the explicit forms for the Yukawa
coupling thresholds, see Ref.\cite{gsth,sth}.

\begin{eqnarray}
  \frac{2\pi}{\alpha^-_i(\Lambda)} \>=\>
  \frac{2\pi}{\alpha^+_i(\Lambda)} - \Delta_i^S(\Lambda) -
  \Delta_i^{DR}\,,
\end{eqnarray}
\begin{eqnarray}
  \Delta_1^S(\Lambda) &\!\! = \!\!& \frac{1}{15} \ln \left
    ( \frac{M_{\tilde{t}_+}}{\Lambda} \right) + \frac{4}{15} \ln \left
    ( \frac{M_{\tilde{t}_-}}{\Lambda} \right) - \frac{1}{5} \sin^2
  \theta_{\tilde t} \,\ln \left
    ( \frac{M_{\tilde{t}_+}}{M_{\tilde{t}_-}} \right) \nonumber
  \\[1mm]
  && - \frac{1}{30} \ln \left( \frac{M_{\tilde{b}_+}}{\Lambda} \right)
  + \frac{1}{15} \ln \left( \frac{M_{\tilde{b}_-}}{\Lambda} \right) -
  \frac{1}{10} \sin^2  \theta_{\tilde b} \ln \left
    ( \frac{M_{\tilde{b}_+}}{M_{\tilde{b}_-}} \right) \nonumber
  \\[1mm]
  && + \frac{1}{10} \ln \left( \frac{M_{\tilde{\tau}_+}}{\Lambda}
  \right) + \frac{1}{5} \ln \left( \frac{M_{\tilde{\tau}_-}}{\Lambda}
  \right) - \frac{1}{10} \sin^2 \theta_{\tilde \tau} \ln \left
    ( \frac{M_{\tilde{\tau}_+}}{M_{\tilde{\tau}_-}} \right) \nonumber
  \\[1mm]
  && + \sum_{i = 1, 2} \left\{ + \frac{1}{15} \ln \left
      ( \frac{M_{\tilde{u}_{L i}}}{\Lambda} \right) + \frac{4}{15} \ln
    \left( \frac{M_{\tilde{u}_{R i}}}{\Lambda} \right) - \frac{1}{30}
    \ln \left( \frac{M_{\tilde{d}_{L i}}}{\Lambda} \right)
  \right. \nonumber \\[1mm]
  && \left. + \frac{1}{15} \ln \left( \frac{M_{\tilde{d}_{R
            i}}}{\Lambda} \right) + \frac{1}{10} \ln \left
      ( \frac{M_{\tilde{e}_{L i}}}{\Lambda} \right) + \frac{1}{5} \ln
    \left( \frac{M_{\tilde{e}_{R i}}}{\Lambda}
    \right)  \right\} \nonumber \\[1mm]
  && + \frac{2}{5} \ln \left( \frac{m_{\tilde{\chi}_2}}{\Lambda}
  \right) - \frac{1}{5} \left( \sin^2 \theta_L + \sin^2 \theta_R
  \right) \ln \left( \frac{m_{{\tilde \chi}_2}}{m_{{\tilde \chi}_1}}
  \right)\,, \\[1mm]
  \Delta_2^S(\Lambda) &\!\! = \!\!& \frac{1}{3} \ln \left
    ( \frac{M_{\tilde{t}_+}}{\Lambda} \right) - \frac{1}{3} \sin^2
  \theta_{\tilde t} \,\ln \left
    ( \frac{M_{\tilde{t}_+}}{M_{\tilde{t}_-}} \right) + \frac{1}{3}
  \sum_{i = 1, 2} \ln \left( \frac{M_{\tilde{u}_{L i}}}{\Lambda}
  \right) \nonumber \\[1mm]
  && + \frac{1}{6} \ln \left( \frac{M_{\tilde{b}_+}}{\Lambda} \right)
  - \frac{1}{6} \sin^2 \theta_{\tilde b} \,\ln \left
    ( \frac{M_{\tilde{b}_+}}{M_{\tilde{b}_-}} \right) + \frac{1}{6}
  \sum_{i = 1, 2} \ln \left( \frac{M_{\tilde{d}_{L i}}}{\Lambda}
  \right) \nonumber \\[1mm]
  && + \frac{1}{6} \ln \left( \frac{M_{\tilde{\tau}_+}}{\Lambda}
  \right) -  \frac{1}{6} \sin^2 \theta_{\tilde \tau} \,\ln \left
    ( \frac{M_{\tilde{\tau}_+}}{M_{\tilde{\tau}_-}} \right) +
  \frac{1}{6} \sum_{i = 1, 2} \ln \left( \frac{M_{\tilde{e}_{L
          i}}}{\Lambda} \right)    \nonumber \\[1mm]
  && + \frac{4}{3} \ln \left( \frac{m_{\tilde{\chi}_1}}{\Lambda}
  \right) + \frac{2}{3} \ln \left( \frac{m_{\tilde{\chi}_2}}{\Lambda}
  \right) + \frac{1}{3} \left( \sin^2 \theta_L + \sin^2 \theta_R
  \right) \ln \left( \frac{m_{{\tilde \chi}_2}}{m_{{\tilde \chi}_1}}
  \right)\,, \\[1mm]
  \Delta_3^S(\Lambda) &\!\! = \!\!& 2 \ln \left
    ( \frac{M_{\lambda_3}}{\Lambda} \right) + \frac{1}{6} \sum_{i =
    \pm} \left\{ \ln \left( \frac{M_{\tilde{t}_i}}{\Lambda} \right) +
    \ln \left( \frac{M_{\tilde{b}_i}}{\Lambda} \right) \right\}
  \nonumber \\[1mm]
  && + \frac{1}{6} \sum_{i = 1, 2} \left\{ \ln \left
      ( \frac{M_{\tilde{u}_{L i}}}{\Lambda} \right) + \ln \left
      ( \frac{M_{\tilde{u}_{R i}}}{\Lambda} \right)+ \ln \left
      ( \frac{M_{\tilde{d}_{L i}}}{\Lambda} \right) + \ln
      \left( \frac{M_{\tilde{d}_{R i}}}{\Lambda} \right) \right\},
\end{eqnarray}
\vspace*{-2mm}
\begin{eqnarray}
  \hspace*{-3mm} \Delta_i^{DR} \,=\, -\, \frac{C_2(G_i)}{12 \pi} \,=\,
  \cases{ \displaystyle{- \frac{1}{4 \pi}} \hspace*{7mm}
    ( \;\mbox{for}\; SU(3)_C \,) \vspace*{1.5mm} \cr \vspace*{3mm}
    \displaystyle{- \frac{1}{6 \pi}} \hspace*{7mm} ( \;\mbox{for}\;
    SU(2)_W \,) \cr \hspace*{3mm} 0  \hspace*{10.5mm} ( \;\mbox{for}\;
    U(1)_Y\, ) \cr }
\end{eqnarray}
where the squark and chargino mixing angles are given by
\begin{eqnarray}
  && \hspace*{-2cm} \tan 2 \theta_{\tilde t} \,=\, \frac{2| m_t \left
      ( A_t - \mu \cot \beta \right) |}{ m^2_{{\tilde Q}_3} -
    m^2_{\tilde t} + \left( \frac{1}{2} - \frac{4}{3} \sin^2 \theta_W
    \right) M_Z^2 \cos 2 \beta } \,, \\[1mm]
  && \hspace*{-2cm} \tan 2 \theta_{\tilde b} \,=\, \frac{2| m_b \left
      ( A_b - \mu \tan \beta \right) |}{ m^2_{{\tilde Q}_3} -
    m^2_{\tilde b} - \left( \frac{1}{2} - \frac{2}{3} \sin^2 \theta_W
    \right) M_Z^2 \cos 2 \beta } \,, \\[1mm]
  && \hspace*{-2cm} \tan 2 \theta_{\tilde \tau} \,=\, \frac{2| m_\tau
    \left( A_\tau - \mu \tan \beta \right) |}{ m^2_{{\tilde L}_3} -
    m^2_{\tilde \tau} - \left( \frac{1}{2} - 2 \sin^2 \theta_W \right)
    M_Z^2 \cos 2 \beta } \,,\\[1mm]
  && \hspace*{-2cm} \tan 2 \theta_L \,=\, \frac{ 2\sqrt{2} M_W \left
      ( M_{\lambda_2} \cos \beta + \mu \sin \beta \right)}
  { M_{\lambda_2}^2 - \mu^2 - 2 M_W^2 \cos 2\beta } \,, \\[2mm]
  && \hspace*{-2cm} \tan 2\theta_R \,=\, \frac{ 2\sqrt{2} M_W \left
      ( M_{\lambda_2} \sin \beta + \mu \cos \beta \right)}
  { M_{\lambda_2}^2 - \mu^2 + 2 M_W^2 \cos 2\beta } \,.
\end{eqnarray}

\vspace*{2cm}
\setlength{\baselineskip}{15.5pt}

\end{document}